\documentclass[12pt]{article}
\usepackage{amsmath}
\usepackage{graphicx}
\usepackage{enumerate}
\usepackage{natbib}
\usepackage{url} 

\usepackage{subcaption}
\usepackage{graphicx}
\usepackage[export]{adjustbox}
\usepackage{mathtools}
\usepackage{soul}
\usepackage{lipsum}

\usepackage{mathtools}

\usepackage{bm}
\usepackage{bbm}
\usepackage{float}
\usepackage{tikz}
\usepackage{amsthm}
\usepackage{amssymb}
\usepackage{longtable}
\usepackage{mathtools,cancel}
\usepackage{url}
\usetikzlibrary{positioning}
\usetikzlibrary{graphs} 
\usetikzlibrary[graphs]
\usepackage{multirow}
\usepackage{lscape}
\usepackage[english]{babel}
\usepackage{amsmath}
\usepackage{authblk}
\usepackage{xr}

\usepackage{tikz}
\usetikzlibrary{positioning}

\RequirePackage[OT1]{fontenc}

\newcommand{\vat}{\mathbb{E}}

\usepackage{mathtools}
\newcommand{\defeq}{\vcentcolon=}
\newcommand\ind{\bot\hspace*{-6pt}\bot}

\newcommand{\bc}{\begin{center}}
	\newcommand{\ec}{\end{center}}

\newcommand{\bit}{\begin{itemize}}
	\newcommand{\eit}{\end{itemize}}

\newcommand{\be}{\begin{eqnarray*}}
	\newcommand{\ee}{\end{eqnarray*}}
\newcommand{\ben}{\begin{eqnarray}}
	\newcommand{\een}{\end{eqnarray}}

\newcommand{\g}{\,\vert\,}

\newcommand{\D}{\mathcal{D}}

\newcommand{\pa}{\mathrm{pa}}
\newcommand{\fa}{\mathrm{fa}}

\newcommand{\bn}{\bm{n}}
\newcommand{\bx}{\bm{x}}

\newcommand{\bS}{\bm{S}}
\newcommand{\bX}{\bm{X}}

\usepackage{tikz}
\usetikzlibrary{graphs}
\usetikzlibrary{positioning}

\newcommand{\black}{\color{black}}

\newcommand{\blind}{1}

\usepackage{soul}	

\newtheorem{prop}{Proposition}[section]

\newcommand{\Xj}{\mathcal{X}_j}
\newcommand{\Xpaj}{{\mathcal{X}}_{\pa(j)}}
\newcommand{\Xfaj}{{\mathcal{X}}_{\fa(j)}}
\newcommand{\X}{{\mathcal{X}}}
\newcommand{\thetajpaj}{\theta^{j \g \pa(j)}_{m \g s}}

\newcommand{\thetaj}{\boldsymbol{\theta}^{j \g \pa(j)}_{s}}
\newcommand{\aj}{\boldsymbol{a}^{j \g \pa(j)}_{s}}
\newcommand{\ajpaj}{a^{j \g \pa(j)}_{m \g s}}
\newcommand{\xl}{\boldsymbol{x}^{(l)}}
\newcommand{\nk}{n^{-i}_k}
\newcommand{\xii}{\boldsymbol{x}^{(i)}}
\newcommand{\nfaj}{\prescript{}{k}n^{\fa(j)}_{(\tilde{m}_j, \tilde{s}_j)}}
\newcommand{\npaj}{\prescript{}{k}n^{\pa(j)}_{\tilde{s}_j}}
\newcommand{\mj}{\tilde{m}_j}
\newcommand{\sj}{\tilde{s}_j}

\newcommand{\btheta}{\boldsymbol{\theta}}

\usepackage{bm}
\usepackage{bbm}
\usepackage{float}
\usepackage{tikz}
\usepackage{amsthm}
\usepackage{amssymb}
\usepackage{longtable}
\usepackage{mathtools,cancel}
\usepackage{url}
\usetikzlibrary{positioning}
\usetikzlibrary{graphs} 
\usetikzlibrary[graphs]
\usepackage{multirow}
\usepackage{lscape}
\usepackage[english]{babel}
\usepackage{amsmath}
\usepackage{authblk}
\usepackage{xr}

\usepackage{tikz}
\usetikzlibrary{positioning}

\RequirePackage[OT1]{fontenc}

\newcommand{\baa}{\begin{eqnarray}}
	\newcommand{\eaa}{\end{eqnarray}}

\usepackage{xr}

\date{}

\begin{document}

\def\spacingset#1{\renewcommand{\baselinestretch}%
{#1}\small\normalsize} \spacingset{1}


\if1\blind
{
  \title{\bf Bayesian nonparametric mixtures of categorical directed graphs for heterogeneous causal inference}
  \author{Federico Castelletti and Laura Ferrini \\
    Department of Statistical Sciences, Universit\`{a} Cattolica del Sacro Cuore, Milan}
  \maketitle
} \fi

\if0\blind
{
  \bigskip
  \bigskip
  \bigskip
  \begin{center}
    {\LARGE\bf Bayesian nonparametric mixtures of categorical directed graphs for heterogeneous causal inference}
\end{center}
  \medskip
} \fi

\bigskip
\begin{abstract}
Quantifying causal effects of exposures on outcomes, such as a treatment and a disease respectively, is a crucial issue in medical science for the 
administration of effective therapies.
Importantly, any related causal analysis should account for all those variables, e.g.~clinical features,
that 
can act as risk factors involved in the occurrence of a disease.
In addition, the selection of targeted strategies for therapy administration requires to quantify such treatment effects at personalized
level rather than at population level.
We address these issues by proposing a methodology based on categorical Directed Acyclic Graphs (DAGs) which provide an effective tool to infer
causal relationships and causal effects between variables.
We account for population heterogeneity by considering a Dirichlet Process mixture of categorical DAGs, which clusters individuals into homogeneous groups characterized by common causal structures, dependence parameters and causal effects.
We develop computational strategies for Bayesian posterior inference, from which a battery of causal effects at subject-specific level is recovered.
Our methodology is evaluated through simulations and applied to a dataset of breast cancer patients to investigate cardiotoxic side effects that can be induced by the administrated anticancer therapies.
\end{abstract}

\noindent%
\vfill

\newpage
\spacingset{1.9} 

\section{Introduction}

\subsection{Motivation and framework}

Estimating cause-and-effect relations between variables is a pervasive issue in many applied domains and primarily medical science. Typically in this setting, interest lies in measuring the (direct or indirect) effect of a therapy 
on the progression of a disease. Our methodology is motivated by a dataset of patients diagnosed with breast cancer and treated with different oncological therapies. In this context, the protein Human Epidermal growth factor Receptor 2 (HER2) has been identified as one of the main responsibles of tumor progression and growth. Recent studies have shown that therapies targeting HER2 have a strong antitumor effect, improving the overall and progression-free survival. These therapies are commonly based on both monoclonal antibodies, such as trastuzumab, as well as anticancer drugs, in particular antracyclines; see \cite{med1} and \cite{med2}. However, they can cause cardiotoxicity as a side effect, with consequent heart failure and loss of left ventricular contractile function \citep{med3,med4}. Establishing the (causal) effect of anti-HER2 therapies on cardiotoxicity is therefore of key importance for the administration of appropriate anticancer treatments, and to develop strategies for preventing and detecting cardiotoxiciy in high-risk patients. In addition, there exist several factors, such as advanced age, hypertension, valvulopathy and arrhythmia, that predispose to cardiotoxicity; see in particular \cite{med5}, \cite{med6} and references therein. Accordingly, in a related causal-effect analysis one should account for all those clinical features that may act as risk factors in the occurrence of cardiotoxicity.


When several variables are entertained, as in the framework above, one should account for possible interactions/dependencies between them in order to provide a coherent quantification of causal effects. Typically however, such dependence structure is unknown, or can be partially drawn only based on experts' knowledge and one need to learn it from the data.
Graphical models based on Directed Acyclic Graphs (DAGs) offer a powerful tool for this \emph{structure learning} task.
Importantly to our purposes, DAGs allow to properly define the causal effect on a target variable of interest induced by a \emph{hypothetical} intervention on another variable in the system. Additionally, learning such causal effect can be achieved from observational data alone, under suitable causal assumptions on the data generating process \citep{Pearl:2000}.
An important issue in this general framework is however represented by \emph{heterogeneity}, which implies that causal effects
may vary across individuals, as the consequence of an existing, yet unknown, clustering structure in the population.
Causal-inference methodologies accounting for heterogeneity can provide a more reliable quantification of treatment effects across patients, leading to personalized strategies for the administration of therapies; see in particular \citet{Stingo:et:al:2015} for an overview.

\subsection{Related work}



Available methods for clustering multivariate (categorical) data include distance- and model-based approaches.
Among the first, \textit{k-modes} \citep{kmodes} is the most popular methodology, which is based on a
modified version of the \textit{k-means} algorithm \citep{MacQueen:1967}
and implements a dissimilarity measure between modes of categorical variables computed across clusters.
On the other side,
poLCA (polytomous variable Latent Class Analysis) \citep{linzer} is a model-based method which applies to a collection of categorical random variables.
In the model, each mixture component corresponds to a multivariate categorical distribution built under the assumption of independence between marginal distributions and for a known number of components in the mixture.
Importantly however, none of these methods accounts for possible dependence relationships between variables in the underlying multivariate statistical model, which instead represents a peculiar feature of our methodology.
Model-based clustering methods have been extensively developed in the Bayesian literature from both a finite and infinite mixture-model perspective; see in particular
\citet{Fruhwirth:Schnatter:et:al:2021}, \citet{Argiento:De:Iorio:2022} and references therein for a review and connections between the two approaches.
In a multivariate Gaussian framework, infinite-mixture models based on a Dirichlet Process (DP) prior are considered by
\citet{Rodriguez:et:al:2009} and \citet{castellettigaussian} for clustering and structure learning of undirected and directed graphs respectively.
Recently, \cite{argientopaci} proposed a finite-mixture model specifically designed for multivariate unordered categorical data. This is based on a newly-introduced class of Hamming distributions which is assumed for each categorical variable marginally,
and from which a joint distribution over $q$ variables is built under the assumption of (local) independence.
Finally, \citet{Malsiner:et:al:2024} proposed a two-layer mixture model which also allows for associations among categorical variables within each mixture-component. Such dependencies arise from a second-layer mixture, assumed within each component of the main mixture model, rather than a multivariate model with allied dependence parameter, which is instead a distinctive feature of our method for causal discovery and inference.



The literature on heterogeneous causal inference has grown extensively in the last years, particularly under the potential outcome framework \citep{Rubin:2005}.
Assuming the existence of latent sub-groups of individuals in the population,
this issue is addressed through the definition and estimation of group-specific causal effects, known as Conditional Average Treatment Effects (CATEs).
Machine learning methods
are employed to identify the underlying clustering structure by \textit{stratification} of subjects based on the levels of available covariates; see \cite{Dominici:Mealli:2021} for a review, \cite{Imbens:et:al:2020}, \cite{Hahn:et:al:2020} and \citet{Bargagli:Stoffi:et:al:2022}, the latter proposing a method which is specifically designed to handle imperfect compliance. Recent methods also aim at improving interpretability of causal results. An instance in this direction is the Causal Rule Ensamble (CRE) method \citep{Bargagli:Stoffi:2021:CRE}, which adopts multiple trees to identify patterns of heterogeneity in the data and to ensure stability in sub-group identification.
Other approaches to heterogeneous causal inference using counterfactuals are based on Bayesian nonparametric methods; see for instance \cite{Linero:Antonelli:2022} and references therein. Among these, \cite{Zorzetto:Canale:et:al:2024} employ a Dependent Probit Stick-Breaking mixture model to simultaneously impute the missing outcomes (counterfactuals), and to identify mutually exclusive groups, thus allowing to estimate causal effects in the presence of population heterogeneity.
Still in a potential outcome framework,
\citet{Roy:et:al:2016} adopt marginal structural models and implement a dependent Dirichlet Process (DP) prior for the evaluation of heterogeneous causal effects of treatments on survival outcomes; \citet{Oganisian:et:al:2021} instead consider a DP mixture of zero inflated regression models for pathological data exhibiting excesses of zeros.
DP priors for clustering and heterogeneous causal inference are also employed by \citet{castellettigaussian} in a multivariate framework based on Gaussian graphical models where causal effects are defined and estimated according to do-calculus theory \citep{Pearl:2000}.









\subsection{Contribution and structure of the paper}

We propose a Bayesian methodology based on a infinite mixture of categorical DAGs for causal discovery and causal effect estimation in the presence of heterogeneous data.
Our model allows for the presence of latent sub-groups of individual/patients in the sample, each characterized by a possibly different causal structure and battery of related causal-effect parameters.
Specifically, we assume that the multivariate distribution of the observables belongs to a Dirichlet Process (DP) mixture of categorical DAG models. Each mixture component reflects a factorization of the sampling distribution satisfying a set of conditional independencies imposed by the DAG. Under the latter, causal effects between variables are then defined according to do-calculus.
With regard to the DP prior, we define a baseline measure over the space of priors on $(\D,\btheta)$ where $\D$ is a DAG and $\btheta$ the parameter of a categorical DAG model.
We employ a constructive procedure based on local and global parameter independence to assign priors to DAG parameters, providing closed-form expressions for both the
prior and posterior predictive of DAGs,
as well as for the posterior distribution of DAG-parameters.
We then leverage these results to develop a computational scheme for posterior inference of our DP model.
When applied to breast cancer data, our methodology ultimately allows to quantify treatment effects of assigned therapies w.r.t.~the occurrence of cardiotoxicity at subject-specific levels, thus leading to a more reliable decision process for the development of personalized therapies.

The rest of the paper is organized as follows.
In Section \ref{sec:background} we provide some background material on DAGs and causal effects within a categorical modelling framework.
In Section \ref{sec:mixture:categorical:DAGs} we introduce our mixture model based on a DP prior, for which we detail the construction of the baseline mixing measure over the space of DAGs and allied parameters.
We then describe a Markov Chain Monte Carlo (MCMC) strategy for posterior inference in Section \ref{sec:posterior:inference}. In Section \ref{sec:simulations} we evaluate our methodology relative to the tasks of clustering and causal discovery through extensive simulation studies, which include comparisons with alternative state-of-the-art methods.
Section \ref{sec:application} is devoted to the analysis of breast cancer data and includes our causal-effect analysis to evaluate heterogeneous side effects of anti-HER2 therapies with respect to the occurrence of cardiotoxicity. We finally provide a discussion to our methodology in Section \ref{sec:discussion}, together with possible future developments.
Some technical results, including the computation of prior and posterior predictive distributions required by our posterior sampler, are reported in the Supplementary Material.


\section{Background}
\label{sec:background}

\subsection{Categorical DAG models}
Consider a Directed Acyclic Graph (DAG) $\D = (V, E)$ with set of nodes $V = \{1, \dots, q\}$ and set of directed edges $E \subseteq V \times V$. For a given $\D$, if $(u,v) \in E$, we say that $u$ is a \emph{parent} of $v$ and let $\pa_{\D}(v)$ be the set of all parents of $v$ in $\D$. Moreover, we let $\fa_{\D}(v) := v \cup \pa_{\D}(v)$ be the \emph{family} of node $v$ in the DAG.
Consider now a collection of random variables $X = (X_1, \dots, X_q)$, such as clinical features that can be measured on patients, and binary categorical variables indicating the administration of a therapy and the absence/presence of a disease. In the following, we assume that
each $X_j$, $j\in V$, is categorical with set of levels $\Xj$ and let $x_j\in\Xj$ be one of its levels. Accordingly, $X \in \X : = \times_{j \in V} \Xj$, whose generic element is $x\in\X$.
In addition, if for any $S\subset V$
we let $X_S=(X_j,j\in S)$, then $X_S \in \X : = \times_{j \in S} \Xj$, with $x_S\in \X_S$.
Under $\D$, the joint probability $p(x)=\text{Pr}(X_1 = x_1, \dots, X_q = x_q)$
admits the factorization
\begin{equation} \label{eq1:jointprob}
	p(x) = \prod_{j =1}^q \Pr(X_j = x_j\g X_{\pa(j)} = x_{\pa(j)}).  
\end{equation}
For the remainder of this section we omit DAG $\D$ from our notation and reason conditionally on a fixed DAG.
Let now $\theta_{s}^{S}=\Pr(X_S = s)$, $s \in \X_S$, be a marginal probability for variables in $S\subseteq V$.
Moreover, let $\thetajpaj = \Pr(X_j = m \g X_{\pa(j)} = s)$ be
a conditional probability for $X_j$ given configuration (level) $s$ of $X_{\pa(j)}$, with $m\in \X_j, s \in \X_{\pa(j)}$.
Consider $n$ observations from $X$, $\boldsymbol{x}^{(1)}, \dots, \boldsymbol{x}^{(n)}$, where each $\boldsymbol{x}^{(i)} = (x^{(i)}_1, \dots, x^{(i)}_q)^\top$, and $\boldsymbol{x}^{(i)} \in \X$, $i = 1, \dots, n$. Also, let $\bx_S^{(i)}$ be the sub-vector of $\bx^{(i)}$ with components indexed by $S\subset V$.
If we collect the $\boldsymbol{x}^{(i)}$'s into an $(n, q)$ data matrix $\bX$, then the likelihood function can be written as
\begin{equation}
	\label{eq2:likelihood}
	\begin{aligned}
		p(\bX \g \btheta)
		&=
		\prod_{i = 1}^n \left\{ \prod_{x \in \X} \left\{p(\bx^{(i)} \g \btheta) \right\}^{\mathbbm{1} \{\bx^{(i)} = x \} }\right\} \\
		&= \prod_{j = 1}^q \left\{ \prod_{s \in \X_{\pa(j)}} \left\{ \prod_{m \in \Xj} \left\{\thetajpaj\right\}^{n^{\fa(j)}_{(m,s)}}\right\} \right\},
	\end{aligned}
\end{equation}
now emphasizing the dependence on the DAG-parameter $\btheta$ (corresponding to the collection of conditional probabilities in the equation) and where
$
n^{\fa(j)}_{(m,s)} = \sum_{i = 1}^n \mathbbm{1} \left\{\boldsymbol{x}^{(i)}_{\fa(j)} = (m,s) \right\}
$
is the number of observations for which $X_{\fa(j)} = (m,s)$. See also \citet{castelletti2023joint} for further notation on categorical DAG models.

\subsection{Causal effects for categorical DAGs} \label{causal_eff}


For a given collection of random variables whose multivariate distribution factorizes according to a DAG, we now focus on
the \textit{causal effect} of
an intervention on $X_h$, $h\in V$, on a response variable of interest, say $X_j := Y$, $j\ne h$.
In practice, such an intervention corresponds to
assigning a treatment to an individual, equivalently fixing $X_h =\tilde{x}$, where $X_h$ is typically an \emph{exposure} of $Y$, and this action can be denoted using Pearl's do-operator $\textnormal{do}(X_h = \tilde{x})$ \citep{pearl1}.
This implies a change in the \emph{observational} distribution \eqref{eq1:jointprob}, leading to the so-called \emph{post-intervention} distribution
\begin{equation}
	\label{trunc}
	p(x \g \textnormal{do}(X_h = \tilde{x})) =
	\begin{cases}
		\prod\limits_{j \neq h} p\big(X_j = x_j\g X_{\pa(j)}=x_{\pa(j)}\big) & \textrm{if} \ X_h = \tilde{x} \\
		\,\,0 & \textrm{otherwise}.
	\end{cases}
\end{equation}
Assuming for simplicity that both $X_h$ and $Y$ are binary variables with levels in $\{0, 1\}$, the \emph{causal effect} of
$\textnormal{do}\{X_h = \tilde{x} \}$ on $Y$
can be defined as
\begin{equation}
	\label{eq:causal:effect:definition}
	c_{y,h} \defeq \vat\big[Y \g \textnormal{do}(X_h = 1)\big]- \vat\big[Y \g \textnormal{do}(X_h = 0)\big];
\end{equation}
see \cite{pearl1}.
More in general, if $X_h$ is polytomous with levels labeled as $\{0,1,\dots,L\}$, one can define a battery of causal effects by considering $X_h = l$, for each $l=1,\dots,L$ in the first expectation of Equation \eqref{eq:causal:effect:definition}.
According to the definition above, $c_{y,h}$ involves a (marginal) post-intervention distribution of $Y$. However, because of \eqref{trunc}, the latter can be expressed in terms of observational distributions, simply by conditioning and then marginalizing w.r.t.~a \emph{valid adjustment set} $Z \subset X$; see \cite{pearl1}. A common choice for such an adjustment set is $Z = X_{\pa(h)}$, namely the parents of $X_h$, leading to
\begin{equation}
	\begin{aligned}
		\label{eq:cv}
		c_{y,h} =
		\sum_{s \in \X_{\pa(h)}} \vat\big(Y\g & X_h=1, X_{\pa(h)}=s\big)
		\Pr\big(X_{\pa(h)}=s\big)\\
		-&
		\sum_{s \in \X_{\pa(h)}} \vat\big(Y\g X_h=0, X_{\pa(h)}=s\big)
		\Pr\big(X_{\pa(h)}=s\big);
	\end{aligned}
\end{equation}
see in particular Theorem 3.2.3 in \cite{Pearl:2000}.
Under model \eqref{eq2:likelihood}, the causal effect in \eqref{eq:cv} can be expressed as a function of the DAG parameter $\btheta$ as
\begin{equation} \label{causaleffect}
	\begin{aligned}
		\gamma_{y,h}(\boldsymbol{\theta}) = \sum_{s \in \mathcal{X}_{\pa(h)}}  \left\{\left( \theta^{Y\g \fa(h)}_{1\g(1,s)} - \theta^{Y\g \fa(h)}_{1|(0,s)} \right) \theta^{\pa(h)}_s \right\}.
	\end{aligned}
\end{equation}


\section{DP mixture of categorical DAG models}
\label{sec:mixture:categorical:DAGs}


In this section we introduce our Dirichlet Process (DP) mixture of categorical DAG models. This can be written using the following hierarchical structure
\begin{equation} \label{dpmixture}
	\begin{aligned}
		\boldsymbol{x}^{(i)} \g \boldsymbol{\theta}_i , \D_i &\sim p(\boldsymbol{x}^{(i)} \g \boldsymbol{\theta}_i, \D_i) \\
		(\boldsymbol{\theta}_i, \D_i) \g H &\sim H \\
		H &\sim DP(M_0, \alpha)
	\end{aligned}
\end{equation}
where $DP(M_0, \alpha)$ denotes a DP prior with baseline $M_0$ and concentration parameter $\alpha$ \citep{Ferguson:1973}, and we now emphasize the dependence on DAG $\D_i$ in the sampling distribution $p(\bx^{(i)}\g\btheta_i,\D_i)$.

A property of model \eqref{dpmixture} is that it induces a partition of the observations $\boldsymbol{x}^{(1)}, \dots, \boldsymbol{x}^{(n)}$ into clusters, with individuals assigned to the same cluster sharing the same DAG $\D$ and DAG parameter $\btheta$. Moreover, the expected number of clusters is controlled by $\alpha$: each observation $\boldsymbol{x}^{(i)}$ is associated with its own $(\boldsymbol{\theta}_i, \D_i)$-parameter as $\alpha \rightarrow \infty$; on the contrary, if $\alpha \rightarrow 0$, then all observations are assigned to the same cluster, leading to a standard categorical DAG model \citep{castelletti2023joint}; see also \cite{muller2013dirichlet} for related properties of the DP prior.

Let now $K \leq n$ be the number of unique values among $(\boldsymbol{\theta}_1, \D_1), \dots, (\boldsymbol{\theta}_n, \D_n)$, and $\{\xi_i\}_{i=1}^n$ a sequence of (cluster) indicator variables such that $\xi_i \in \{1, \dots, K\}$ and $(\boldsymbol{\theta}_i, \D_i) = (\boldsymbol{\theta}_{\xi_i}, \D_{\xi_i})$.
Conditionally on $\{\xi_i\}_{i=1}^n$, observations are i.i.d.~within each cluster, so that the likelihood can be written as
\begin{equation} \label{eq:lik_clustering}
	\begin{aligned}
		p\left(\bX \g \{\xi_i\}_{i=1}^n, \{\btheta_i\}_{i=1}^n, \{\D_i\}_{i=1}^n\right) &= \prod_{k = 1}^K  \left\{ \prod_{i : \xi_i = k} p\left(\boldsymbol{x}^{(i)} \g \btheta_{\xi_i}, \D_{\xi_i}\right) \right\} \\
		&= \prod_{k=1}^K p\big(\bX^{(k)}\g\btheta_k,\D_k\big),
	\end{aligned}
\end{equation}
with $p\big(\bX^{(k)}\g\btheta_k,\D_k\big)$ as in Equation \eqref{eq2:likelihood}, and where $\bX^{(k)}$ is the $(n_k,q)$ matrix collecting all observations $\bx^{(i)}$ such that $\xi_i=k$.
Additionally, a generic count involved in the $k$-th component above will be denoted as $\prescript{}{k}n^{\fa(j)}_{(m,s)}=\sum_{i:\xi_i=k}\mathbbm{1}\big\{\bx^{(i)}_{\fa(j)}=(m,s)\big\}$, which corresponds to the number of observations in cluster $k$ for which the level taken by variables $X_{\fa(j)}$ is equal to $(m,s)$.
An alternative representation of the DP prior
is based on the so-called \textit{stick-breaking} process \citep{Sethuraman:1994}.
Accordingly, $H$ can be written in the form
\begin{equation}
	H = \sum_{k = 1}^{\infty}\omega_k \delta_{(\boldsymbol{\theta}_k, \D_k)}
\end{equation}
where $\delta_{(\boldsymbol{\theta}_k, \D_k)}$ is a degenerate probability measure placing all of its mass on $\left\{\boldsymbol{\theta}_k, \D_k\right\}$ and $\left\{\btheta_k, \D_k\right\}_{k = 1}^{\infty} \overset{\textnormal{iid}}{\sim} M_0$. Moreover, the weights $\{\omega_k\}_{k=1}^{\infty}$ satisfy $\omega_1 = v_1$, and $\omega_k = v_k \prod_{h < k} (1-v_h)$, where $\{v_k\}_{k = 1}^\infty \overset{\textnormal{iid}}{\sim} \textnormal{Beta}(1, \alpha)$, with $\alpha$ the concentration parameter of the DP prior.
In the following we will assign $\alpha \sim \textnormal{Gamma}(c,d)$ following \cite{escobar}.

In the next sections we detail the construction of the baseline $M_0$.
This is structured as $M_0=p(\btheta\g\D)p(\D)$, where the former term corresponds to a prior on the DAG parameter $\btheta$ conditionally on DAG $\D$, while the latter is a marginal prior over DAGs.

\subsection{Baseline on DAG parameter}
\label{sec:baseline:theta}

Conditionally on DAG $\D$, we first assign a prior $p(\btheta\g\D)$.
To this end, consider for each node $j \in \{1,\dots,q\}$ and $s \in \X_{\pa(j)}$ the parameter
$
\big(\thetajpaj, m \in \Xj\big) := \thetaj
$
corresponding to a $|\Xj|$-dimensional vector collecting conditional probabilities for variable $X_j$, given a configuration $s$ of its parents $X_{\pa(j)}$.
We assign to each $\thetaj$ a Dirichlet prior with hyper-parameter $\aj= \big(a^{j\g\pa(j)}_{m\g s}, m \in \Xj\big)$,
written as $\thetaj \sim \textnormal{Dirichlet}(\aj)$, whose p.d.f. is
\begin{equation}
	\label{dirichlet:prior}
	\begin{aligned}
		p\big(\thetaj\big) = 
		h\big(\aj\big) \prod_{m \in \Xj} \left\{ \thetajpaj \right\}^{\ajpaj - 1},
	\end{aligned}
\end{equation}
and where $h\big(\aj\big)$ is the prior normalizing constant.
Let now $\btheta^{j\g\pa(j)}=\big(\thetaj, s \in \X_{\pa(j)}\big)$.
By assuming
global and local parameter independence \citep{g&l}, respectively $\ind_{j} \btheta^{j\g\pa(j)}$ and $\ind_{s} \thetaj$, a joint prior on $\boldsymbol{\theta} = \big\{\btheta^{j\g\pa(j)}, j \in V \big\}$ can be written as 
\begin{equation}
	\label{prior:theta}
	\begin{aligned}
		p(\boldsymbol{\theta}) = \prod_{j=1}^q \left\{ \prod_{s \in \Xpaj} p\big(\thetaj\big) \right\}.
	\end{aligned}
\end{equation}
In what follows we implement the default choice $\ajpaj = a/|\Xfaj|$, $a>0$, leading to the Bayesian Dirichlet Equivalent uniform (BDEu) score \citep{Heckerman:et:al:1995}, which guarantees that Markov equivalent DAGs are assigned the same marginal likelihood; see also \citet{castelletti2023joint}.

Also notice that the resulting prior is conjugate with the likelihood \eqref{eq2:likelihood} since, for generic dataset $\bX$, $\thetaj \g \bX \sim \textnormal{Dirichlet}\big(\aj + \bn_{s}^{\fa(j)}\big)$
with $\bn_{s}^{\fa(j)} = \big(n_{(m,s)}^{\fa(j)}, m \in \Xj\big)$.
Accordingly, the posterior of $\btheta$ is 
\begin{equation}
	\label{posterior:theta}
	\begin{aligned}
		p(\boldsymbol{\theta}\g \bX) = \prod_{j=1}^q \left\{ \prod_{s \in \Xpaj} p\big(\thetaj\g\bX\big) \right\},
	\end{aligned}
\end{equation}
with each term corresponding to a Dirichlet p.d.f., so that direct sampling from $p(\btheta\g \bX)$ is possible.
Finally, under the same prior, a marginal (i.e.~integrated w.r.t.~to $\btheta$) likelihood $m(\bX\g\D)=\int p(\bX\g\btheta,\D)p(\btheta\g\D)\,d\btheta$ is available and admits the factorization
\begin{equation}
	\label{eq:marg:like:dag}
	m(\bX\g\D)=\prod_{j=1}^q  m(\bX_j \g \bX_{\pa(j)}),
\end{equation}
with
\begin{equation}
	m\big(\bX_j \g \bX_{\pa(j)}\big) =
	\prod_{s\in\X_{\pa(j)}} \frac{h\big(\aj\big)}{h\big(\aj + \bn^{\fa(j)}_{s}\big)}
\end{equation}
and where $h\big(\aj + \bn^{\fa(j)}_{s}\big)$ is the posterior normalizing constant. See also the Supplementary Material (Section 1) for full details.

\subsection{Baseline on DAGs}
\label{sec:baseline:dags}

Let $\mathcal{S}_q$ be the (discrete) space of \emph{all} DAGs with $q$ nodes.
Additionally, we can restrict $\mathcal{S}_q$ to a subset of DAGs satisfying some structural constraints, typically edge orientations that can be postulated in advance based on the specific real-data problem. As an instance, in our application to breast cancer data we regard age as an exogenous variable and forbid any incoming edge to it from other variables; conversely, we regard the occurrence of cardiotoxic side effect as a response variable and accordingly forbid any outgoing edge from it.
Each DAG $\D=(V,E)$ in $\mathcal{S}_q$ can be represented through a 0-1 adjacency matrix $\boldsymbol{A}^{\D}$, whose $(u,v)$-element $\boldsymbol{A}^{\D}_{u,v}=1$ if $(u,v) \in E$, $0$ otherwise.
Additionally, let $\bS^{\D}$, be the adjacency matrix of the skeleton of $\D$, namely the undirected graph obtained from $\D$ by disregarding edges orientations.
We assign for each $u>v$, $\bS^{\D}_{u,v} \g \pi \overset{\textnormal{\textnormal{iid}}}{\sim} \textnormal{Ber}(\pi)$
where $\pi$ is a prior probability of edge inclusion. We then assume hierarchically
$\pi \sim \textnormal{Beta}(a,b)$, leading to the integrated prior on DAG $\D$
\begin{equation}
	p(\D) \propto p\big(\bS^{\mathcal{D}}\big) =  \frac{\Gamma(a+b)}{\Gamma(a) \Gamma(b)} \cdot \frac{\Gamma\big(|\bS^{\mathcal{D}}| + a\big) \Gamma\big(q(q-1)/2 - |\bS^{\mathcal{D}}| + b\big)}{\Gamma\big(q(q-1)/2 + a + b\big)}, \nonumber
\end{equation} 
where $|\bS^{\mathcal{D}}|$ the number of non-null elements in $\bS^{\mathcal{D}}$, corresponding to the number of edges in $\D$, and $q(q-1)/2$ is the maximum number of edges in a DAG having $q$ nodes.
Sampling from the baseline over DAGs, as required by our MCMC sampler (Section \ref{sec:posterior:inference}), is possible through an acceptance-rejection algorithm over the space $\mathcal{S}_q$; see also the Supplementary Material (Section 2).

\section{Posterior inference}
\label{sec:posterior:inference}

In this section we detail our Markov Chain Monte Carlo (MCMC) strategy for posterior inference of a DP mixture of categorical DAGs.
This is based on a collapsed sampler with DAG parameters integrated out, and which accordingly approximates a marginal posterior over DAGs and cluster indicators $\xi_1,\dots,\xi_n$.
Such output allows for inference about the clustering structure and/or the graphical structures associated with the clusters.
In a second step, DAG parameters can be sampled conditionally on $\xi_1,\dots,\xi_n$ based on Equation \eqref{posterior:theta}.


\subsection{MCMC scheme}

The structure of our baseline measure (Section \ref{sec:baseline:theta}) is such that we can integrate out the DAG parameter $\btheta$, which allows for the implementation of a collapsed sampler approximating the marginal posterior of $\{\xi_i\}_{i = 1}^n, \alpha, \{\D_k\}_{k =1}^K, K$.
The resulting scheme has a Gibbs-sampling structure as Algorithm 2 in \cite{Neal:2000} and implements the following steps.
%
%

\subsubsection{Update of cluster indicators}
\label{update_clusters}

The full conditional of $\xi_i$ is
\begin{equation} \label{prob_cluster}
	\begin{aligned}
		p(\xi_i = k \g \{\xl : l \neq i, \xi_l = k\}, \D_k) \propto
		\begin{cases}
			\nk \ p(\xii \g  \{\xl : l \neq i, \xi_l = k\}, \D_k) & k = 1, \dots, K \\
			\alpha \ p(\xii \g \D_{k}) & k = K + 1,
		\end{cases}
	\end{aligned}
\end{equation}
which corresponds to the probability that subject $i$ is assigned to cluster $k$, conditionally on all the observations currently assigned to that cluster, and on $\D_k$.
In particular, for a non empty cluster $k = 1, \dots, K$, the full conditional is proportional to the product between two terms: the number of observations belonging to cluster $k$ (possibly excluding observation $i$), $\nk = \sum_{l \neq i} \mathbbm{1}\{\xi_l = k\}$, and the posterior predictive distribution $p(\xii \g \{\xl : l \neq i, \xi_l = k\}, \D_k)$ evaluated at $\bx^{(i)}$.
For the latter, we provide in the following proposition a simple closed-form expression.
\begin{prop}[Posterior predictive - non-empty cluster] \label{post_pred_nonempty}
	For a given cluster $k$, consider the data matrix $\bX^{(k)}$ collecting the $n_k$ observations $\big\{\bx^{(l)} : \xi_l=k\big\}$ and an observation $\bx^{(i)}$.
	Then, the posterior predictive of $\bx^{(i)}$ given $\{\xl : l \neq i, \xi_l = k\}$ is
	\begin{equation} \label{posteriorpredictive}
		\begin{aligned}
			p(\xii \g \{\xl : l \neq i, \xi_l = k\}, \D_k) =
			\prod_{j = 1}^q \left\{ \frac{a/|\Xfaj| + \nfaj - \mathbbm{1}\{\xi_i=k\}}{ a/|\Xpaj| + \npaj - \mathbbm{1}\{\xi_i=k\}} \right\} 
		\end{aligned}
	\end{equation}
	where $\mj=\xii_j, \sj=\xii_{\pa(j)}$ and
	$$
	\nfaj=\sum_{l : \xi_l = k} \mathbbm{1}\big\{\xl_{\fa(j)} = (\mj, \sj) \big\}, \quad \npaj = \sum_{l : \xi_l = k} \mathbbm{1}\big\{\xl_{\pa(j)} = \sj\big\}.
	$$
	\begin{proof}
		See Supplementary Material.
	\end{proof}
\end{prop}


The second expression of \eqref{prob_cluster} considers the case of a (new) empty cluster $k = K + 1$, where the DAG $\D_{K+1}$ is sampled from the baseline over $\mathcal{S}_q$.
In such case, the full conditional is proportional to the product of the concentration parameter $\alpha$ and a posterior predictive  which reduces to the marginal likelihood (prior predictive) of a cluster containing subject $i$ only. A related closed-form expression is provided by the following proposition.
\begin{prop}[Posterior predictive - empty cluster] \label{probempty}
	For a new cluster $k = K + 1$, the posterior predictive of $\bx^{(i)}$ coincides with the marginal likelihood and is given by
	\begin{equation}
		p(\xii \g \D_{k}) =  \prod_{j = 1}^q \frac{1}{|\X_j|}.
	\end{equation}
	\begin{proof}
		See Supplementary Material.
	\end{proof}
\end{prop}

\subsubsection{Update of $\alpha$}\label{update_alpha}

Under the DP prior,
the full conditional distribution of $\alpha$ coincides with $p(\alpha\g K)\propto p(K\g \alpha)p(\alpha)$, where in particular
$$
p(K\g\alpha)\propto c_n(K)\alpha^K \frac{\Gamma(\alpha)}{\Gamma(\alpha+n)}
$$
is the prior on the number of clusters induced by the DP
and $c_n(K)$ is a normalizing constant not involving $\alpha$.
Sampling from $p(\alpha\g K)$ can be done by augmenting the distribution through an auxiliary variable $\eta \sim \textnormal{Beta}(1, \alpha)$. It can be shown \citep{escobar}
that under the prior $\alpha \sim \textnormal{Gamma}(c,d)$ the full conditional of $\alpha \g K, \eta$ corresponds to a mixture of Gamma distributions, specifically
$$
\alpha \g \eta, K \sim g \cdot \textnormal{Gamma}(c+K, d-log \eta) + (1-g) \cdot \textnormal{Gamma}(c+K-1, d-log \eta),
$$
where $g/(1-g) =  (c + K -1) / n(d - log \ \eta)$.

\subsubsection{Update of DAGs and sampling of DAG parameters}
\label{update_dags}

Let $K$ be the number of clusters and $\xi_1,\dots,\xi_n$ the cluster indicators, with each $\xi_i\in\{1,\dots,K\}$.
For a given $k\in\{1,\dots,K\}$,
let $\{\bx_i : \xi_i = k\}$ be the set of observations currently assigned to cluster $k$, and $\bX^{(k)}$ the implied $(n_k,q)$ data matrix; see also Equation \eqref{eq:lik_clustering}.
Without loss of generality, consider a generic cluster and omit for simplicity subscripts $k$ from $\D_k$ and $\bX^{(k)}$.
Update of DAG $\D$ is performed through a Metropolis Hastings step
where a DAG $\widetilde{\D}$ is sampled from a proposal distribution $q(\widetilde{\D}\g\D)$ conditionally on a current DAG $\D$ and it is accepted with probability
$\alpha_{\widetilde{\D}}=\min\{1;r_{\widetilde{\D}}\}$
with
\begin{equation}
	r_{\widetilde{\D}}=
	\frac{m_{}(\bX\g \widetilde{\D})}
	{m_{}(\bX\g \D)}
	\cdot\frac{p(\widetilde{\D})}{p(\D)}
	\cdot\frac{q(\D\g\widetilde{\D})}{q(\widetilde{\D}\g\D)},
\end{equation}
and
$m(\bX\g \widetilde{\D})$ as in Equation \eqref{eq:marg:like:dag}; see also the Supplementary Material for full details.

Finally, conditionally on DAGs $\D_1,\dots,\D_K$ and indicators $\xi_1,\dots,\xi_n$, we can sample each DAG parameter $\btheta_k$
based on Equation \eqref{posterior:theta}, corresponding to the posterior of $\btheta_k$ which is available in closed-form as a product of Dirichlet probability functions; see also Section \ref{sec:baseline:theta}.

\black

\subsection{Posterior summaries}
\label{sec:posterior:summaries}

Output of the MCMC scheme is a collection of cluster indicators, DAGs and DAG parameters approximately drawn from the posterior distribution of our DP mixture model.
Starting from such output, we can provide posterior summaries regarding clustering, DAG structures, as well as DAG-model parameters.
Specifically, let
$K^{(s)}$ be the number of clusters at MCMC iteration $s$,
$\xi_i^{(s)}$, $i=1,\dots,n$, $\D_k^{(s)}$ and $\btheta_k^{(s)}$, $k=1,\dots,K^{(s)}$, be the corresponding realizations of the three sets of parameters.
For clustering purposes, we first recover an $(n,n)$ posterior similarity matrix $\boldsymbol{S}$, with $(i,i')$-element $\boldsymbol{S}_{i, i'}$ corresponding to the (estimated) posterior probability that individuals $i$ and $i'$ are assigned to the same cluster, namely
\begin{equation} \label{eq:post:prob:same:cluster}
	\widehat{p}(\xi_i = \xi_{i'} \g \bX) = \frac{1}{S} \sum_{s = 1}^S \mathbbm{1}\left\{\xi_i^{(s)} = \xi_{i'}^{(s)}\right\}.
\end{equation}
A point estimate of the clustering structure, $\widehat{\boldsymbol{c}}$, can be recovered by assigning individuals $i$ and $i'$ to the same cluster if $\widehat{p}(\xi_i = \xi_{i'}\g \bX)$ exceeds a given threshold, say $z = 0.5$. As an alternative, a clustering estimate can be obtained following \cite{wade:2018} as the partition minimizing the expected Variation of Information (VI); see also Section \ref{sec:simulations}.

From the same MCMC output, we can recover for each subject $i$ a $(q,q)$ matrix collecting estimates of the Posterior Probabilities of edge Inclusion (PPIs). For a given subject $i$ and edge $u \rightarrow v, u \neq v$, its PPI is estimated as
\begin{equation} \label{eq:post:prob:edge:inclusion}
	\widehat{p}_i (u \rightarrow v \g \bX) = \frac{1}{S} \sum_{s = 1}^S \mathbbm{1}\left\{u \rightarrow v \in \D^{(s)}_{\xi_i^{(s)}}\right\},
\end{equation}
corresponding to the proportion of DAGs $\D_i^{(s)}$ in the chain containing the directed edge $u \rightarrow v$.
Finally, a graph estimate at subject-specific level, say $\widehat{\D}_i$, can be obtained by including those edges for which $\widehat{p}_i (u \rightarrow v \g \bX) > z$ for $z \in (0,1)$, e.g $z = 0.5$.

Recall now the definition of causal effect $\gamma_{y,h}(\boldsymbol{\theta})$ provided in Equation \eqref{causal_eff} and assume that the intervened variable and response, $X_h$ and $Y$ respectively, are given so that we can omit them from the notation. A Bayesian Model Averaging (BMA) estimate of $\gamma_i$, the subject-specific causal effect, for $i \in \{1,\dots,n\}$, is given by
\begin{equation} \label{eq:bma}
	\widehat{\gamma_i} = \frac{1}{S} \sum_{s = 1}^S \gamma_{i}^{(s)}\left(\boldsymbol{\theta}_{\xi_i^{(s)}}\right). 
\end{equation}
The resulting collection $\{\widehat{\gamma_1},\dots,\widehat{\gamma}_n\}$ provides estimates of causal effects at individual level, which also naturally account for DAG-model uncertainty through BMA.

\section{Simulations}
\label{sec:simulations}

We conduct simulation studies to evaluate the performance of our methodology relative to the tasks of clustering and structure learning and compare it with alternative methods for clustering multivariate categorical data.
We consider settings with $q = 10$ nodes, number of clusters $K = 2$ and sample sizes $n_1=n_2$ that we range in $\{100, 200, 500\}$. We generate the two DAGs $\D_1$ and $\D_2$ independently, so that the two clusters differ in general by the dependence structure among variables, and by fixing a probability of edge inclusion $\pi=0.2$.
The two categorical datasets $\bX^{(1)},\bX^{(2)}$ are built by discretization of latent Gaussian observations
as detailed in the Supplementary Material (Section 4).
Importantly, discretization is based on a collection of thresholds $g_j \in (-\infty,+\infty)$, that we randomly draw from a $\textnormal{Unif}(\hat{z}_{j, \alpha}, \hat{z}_{j, 1-\alpha})$, where $\hat{z}_{j, \alpha}$ denotes the quantile of order $\alpha$ in the empirical distribution of latent variable $Z_j$, independently across $j$ and for each $k=1,2$; see again our Supplementary Material. We consider $\alpha \in \{0.1,0.4\}$ which implies different degrees of similarity among marginal distributions between clusters.
For benchmark methods that do not consider a dependence structure between variables we expect a lower ability in recovering the true clustering when $\alpha = 0.4$, namely when marginal distributions are more similar across clusters. Finally, under each scenario, a collection of $N=40$ multiple ($K=2$) datasets is generated. \black

\subsection{Clustering}

\begin{center}
	\begin{figure}[H]
		\begin{subfigure}{.5\textwidth}
			\centering
			\includegraphics[width=1\linewidth]{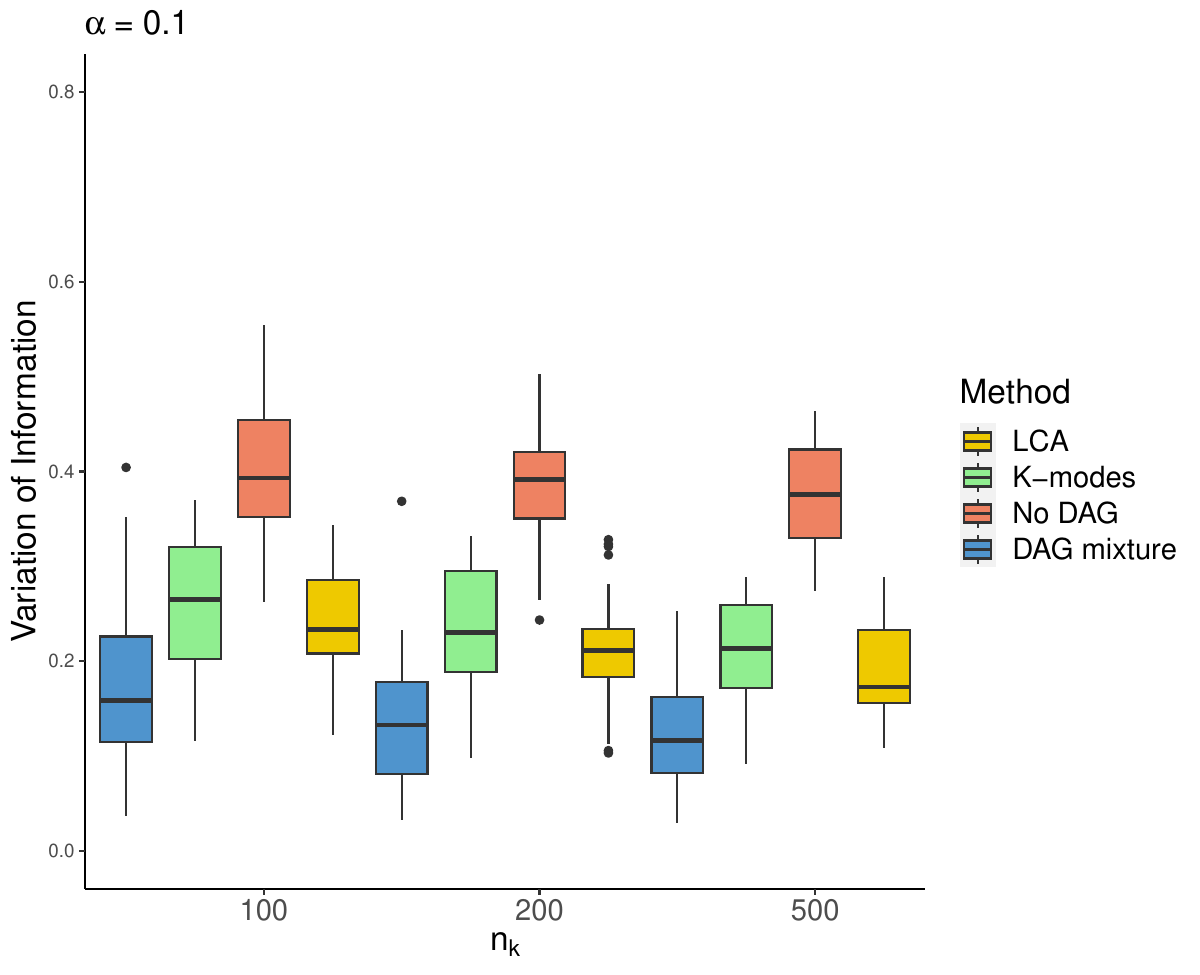}
		\end{subfigure}%
		\begin{subfigure}{.5\textwidth}
			\centering
			\includegraphics[width=1\linewidth]{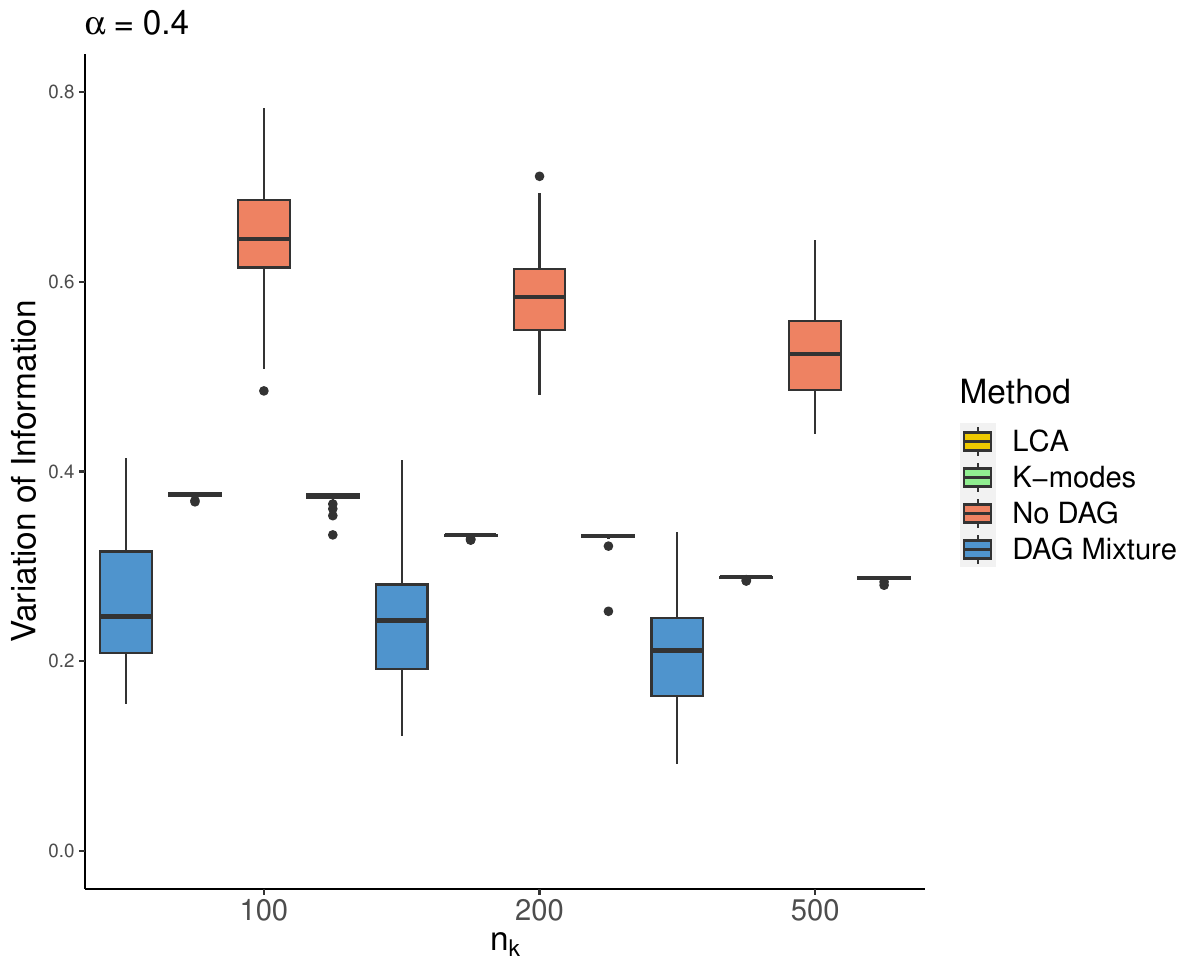}
		\end{subfigure}
		\caption{Simulations. Distribution (across $40$ replicates) of Variation of Information, for the two different simulation scenarios: $\alpha = 0.1$ and $\alpha = 0.4$. Methods under comparison are: Latent Class Model (LCM), K-modes, No DAG, and our DP mixture of DAGs (DAG mixture).}
		\label{fig:clustering_sim} 
	\end{figure}
\end{center}

We evaluate the clustering performance of our method (DAG mixture) w.r.t. state-of-the-art approaches and specifically the Latent Class Model (LCM) 
\citep{goodman,linzer} and K-modes \citep{kmodes}.
For both methods, we input the number of clusters as $K=2$.
Additionally, to emphasize the contribution of a DAG-based model on cluster identification, we also implement a \textit{No DAG} strategy, where for each group the DAG is assumed to be known and corresponds to an empty graph. Performances are assessed by comparing the true partition $\boldsymbol{c}$ with the estimated partitions $\widehat{\boldsymbol{c}}$ based on the Variation of Information (VI). Lower values of the metric correspond to better performances. In addition, we expect scenario with $\alpha = 0.1$ to be characterized by overall better performances than $\alpha = 0.4$, because in the latter the difference between the two clusters is mainly due to the dependency structure among the variables. Results are summarized in the boxplots of Figure \ref{fig:clustering_sim}.
As it appears, all methods tend to improve as the sample size $n_k$ grows, with our DAG mixture model clearly outperforming all the benchmarks under all scenarios.

\subsection{Structure learning}
\begin{center}
	\begin{figure}[H] 
		\begin{subfigure}{.5\textwidth}
			\centering
			\includegraphics[width=1\linewidth]{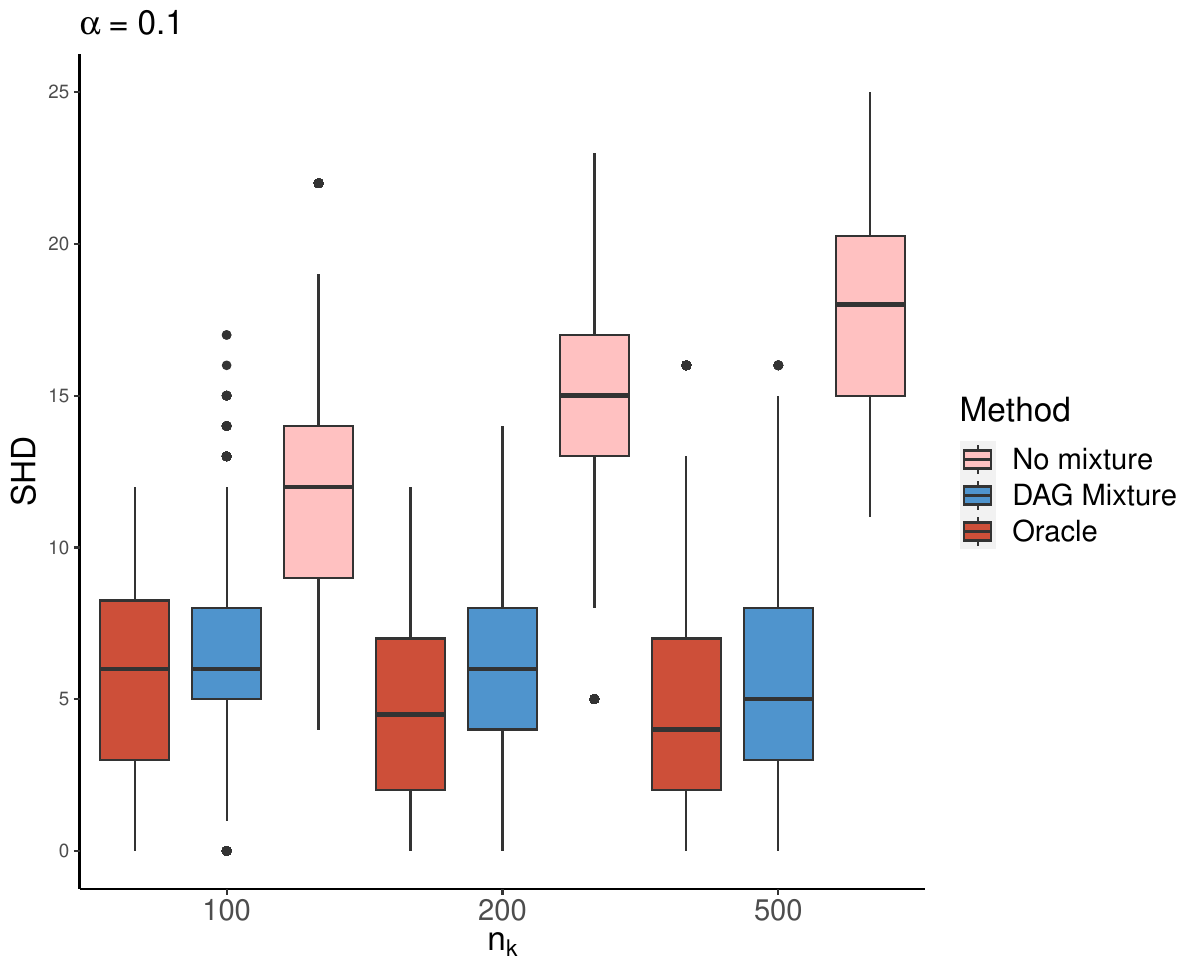}
		\end{subfigure}%
		\begin{subfigure}{.5\textwidth}
			\centering
			\includegraphics[width=1\linewidth]{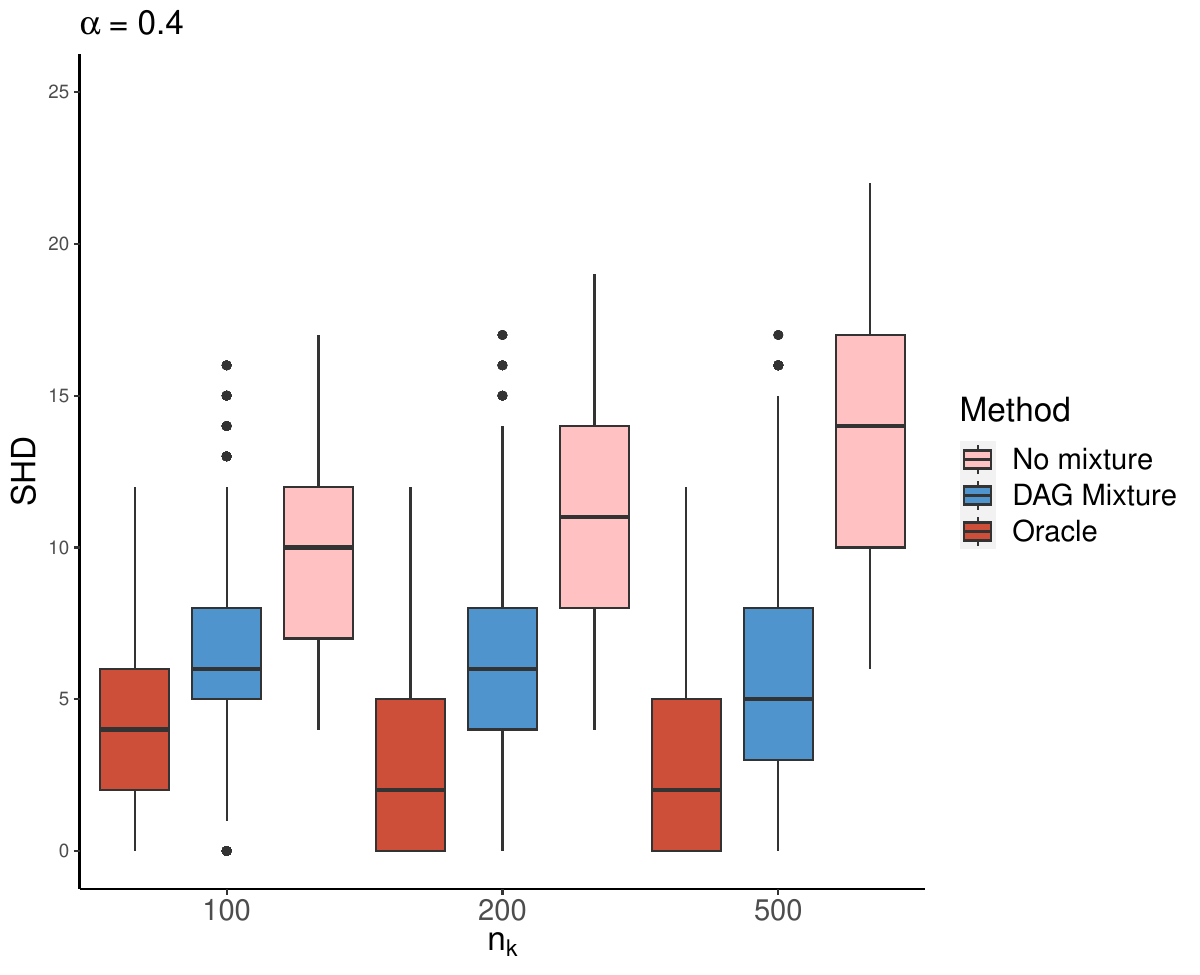}
		\end{subfigure}
		\caption{Simulations. Distribution (across $40$ replicates) of the SHD between estimated DAGs and true DAGs. Methods under comparison are: No mixture, the Oracle version of our method (Oracle) and our DP mixture of DAGs (DAG mixture).}
		\label{fig:shd_res}
	\end{figure}
\end{center}

We now assess the ability of our method in recovering the graphical structure underlying each cluster. To this end, we consider the Structural Hamming Distance (SHD), which represents the number of modifications (edge insertions, edge removals, edge reversal) that are needed to transform the estimated DAG $\widehat{\D}$ into the true DAG $\D$. Specifically, we compare each subject-specific estimated DAG $\widehat{\D}_i, i = 1, \dots, n$ with $\D_{\xi_i}$, where $\xi_i$ is the true class-membership.
In addition, we include the Oracle version of our method, in which the true clustering is assumed to be known, and a ``one-group" naive strategy (No mixture), which instead assigns all subjects to the same cluster and therefore disregard heterogeneity. Results, for each scenario defined by $\alpha$ and $n_k$, are summarized in Figure \ref{fig:shd_res}.
The No mixture strategy, which neglects the clustering structure in the data, performs worse than the other two methods under all scenarios and with a worsen performance as the sample size $n_k$ increases. By converse, the Oracle version of our method performs slightly better than our DP mixture method, a behavior which is more evident under scenario $\alpha=0.4$ where clustering is indeed more difficult. Finally, both methods improve their performance as $n_k$ grows. \black

\section{Analysis of breast cancer data}
\label{sec:application}
\subsection{Dataset and model implementation}
\label{sec:application:dataset}

In this section we analyse a dataset of $n=404$ women diagnosed with HER2+ breast cancer and treated with potentially cardiotoxic 
therapies based on monoclonal antibodies (trastuzumab) and chemotherapy drugs (antracyclines). Variables in the dataset include: demographic and physical features, such as age, height and weight (expressed in terms of Body Mass Index, BMI, and included through a three-level categorical variable); risk factors, such as diagnosis of hypertension (HTA), dyslipidemia (DL), diabetes mellitus (DM), smoking (smoker and ex smoker); past cardiac diseases, namely cardiac insufficiency (CIprev), ischemic cardiomyopathy (ICMprev), arrhythmia (ARRprev), valvulopathy (VALVprev), valve surgery (valvsurgprev). In addition, the dataset provides information regarding treatments, antiHER2 monoclonal therapy (antiHER2) and/or antracyclines (AC),
that were administrated to patients.
Finally, the target variable corresponds to Cancer Therapy-Related Cardiac Dysfunction (CTRCD), a binary outcome indicating the occurrence (1) or not (0) of cardiac dysfunction.
The original dataset is provided as a supplement to \cite{BreastCancerData} and included as supplementary material to our paper.
Notably, all variables are categorical, with the exception of age and heart rate which have been discretized into two dummy variables, corresponding to middle \textit{vs} low, and high \textit{vs} low.
While in general the cardiotoxic effects of the available oncological therapies have been established in the literature, still, the occurrence of CTRCD can vary substantially among patients because of both observed features (such as risk factors) or even unobserved characteristics. Accordingly, it is of interest to quantify
causal effects w.r.t. the occurrence of CTRCD at individual-level, which is crucial for the development and administration of appropriate antiHER2 therapies.

Given the structure of the dataset, we constrain the adjacency matrix of DAGs in such a way that CTRCD can only (potentially) be a response, i.e. no outgoing edges are allowed,
and treat age, BMI, smoker, ex smoker as exogenous variables, i.e.~with no incoming edges from other nodes, while possible links/dependencies between them are allowed.
Moreover, we assume that the absence/presence of risk factors can imply the administration of a therapy (AC, antiHER2), while the converse is not possible.
We implement our mixture model by running the MCMC scheme for $S = 100000$ iterations, which include a burn-in period of $10000$ draws that are discarded from the posterior analysis. With regard to hyperparameters, we fix $c = 3, d = 1$ in the Gamma prior on the DP precision parameter $\alpha$. The common hyperparameter $a$ on the collection of Dirichlet priors on $\thetaj$ is instead fixed as $a = 1$, while in the hierarchical prior on DAGs we fix $a = 1$, $b = 2q$, reflecting an \textit{a priori} assumption of sparsity in the graph space. To assess the convergence of our algorithm we also run two independent MCMC chains; results suggest an overall agreement in terms of clustering (evaluated through posterior similarity matrices), structure learning (based on estimated PPIs) and posterior distribution of causal-effect parameters. Results relative to all such quantities are presented discussed in the following sections.

\subsection{Clustering}
\label{sec:application:clustering}

\begin{center}
	\begin{figure}
		\centering
		\includegraphics[width=.55\linewidth]{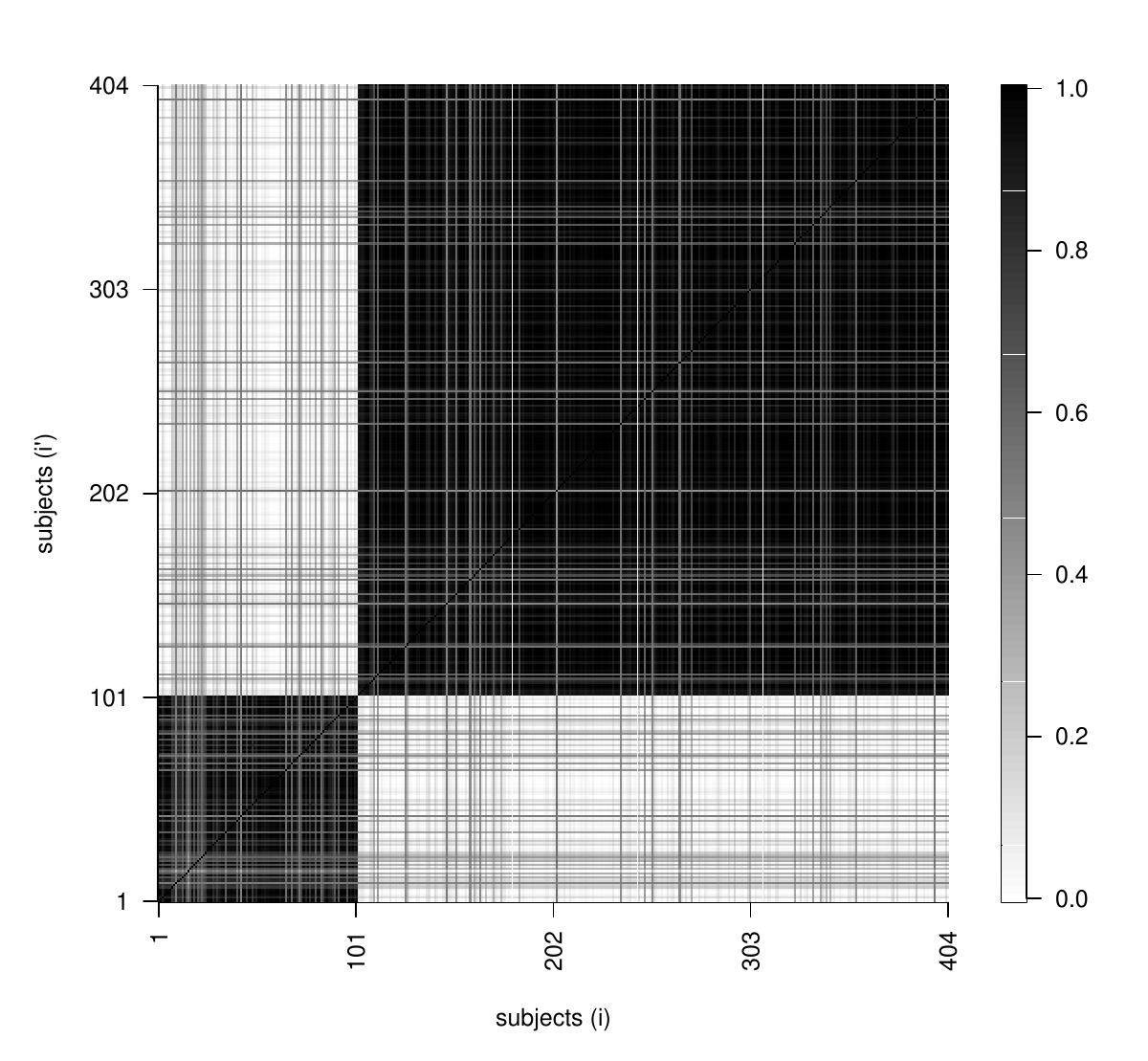} 
		\vspace*{-3mm}
		\caption{Breast cancer data. Estimated posterior similarity matrix, with individuals arranged according to the clustering structure estimated \textit{via} the posterior expectation of VI criterion (cluster labeled as $\widehat{\boldsymbol{c}}_1$ and $\widehat{\boldsymbol{c}}_2$ respectively).}
		\label{application:psm}
	\end{figure}
\end{center}

\begin{center}
	\begin{figure} 
		\begin{subfigure}{.5\textwidth}
			\centering
			\includegraphics[width=1\linewidth]{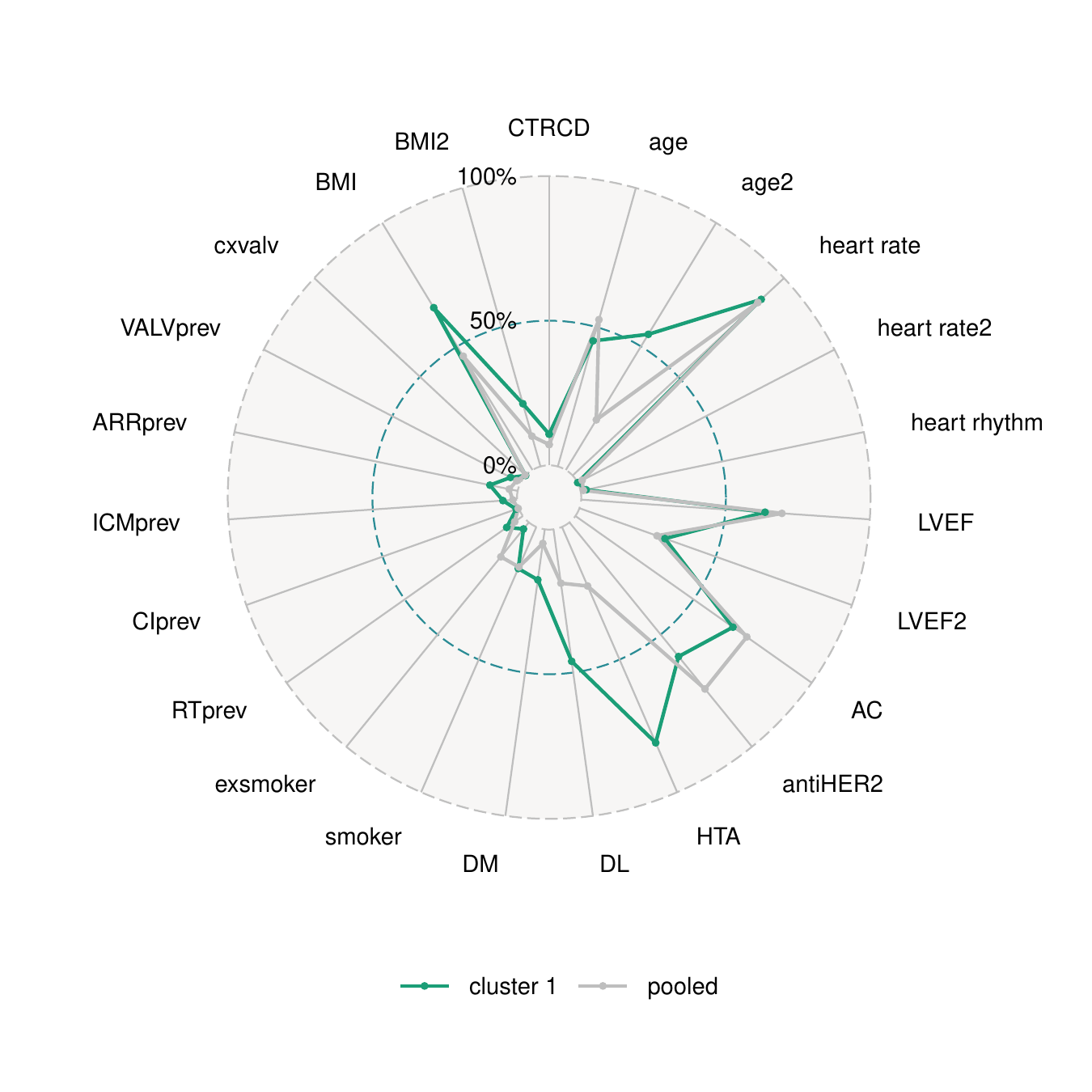}
		\end{subfigure}%
		\begin{subfigure}{.5\textwidth}
			\centering
			\includegraphics[width=1\linewidth]{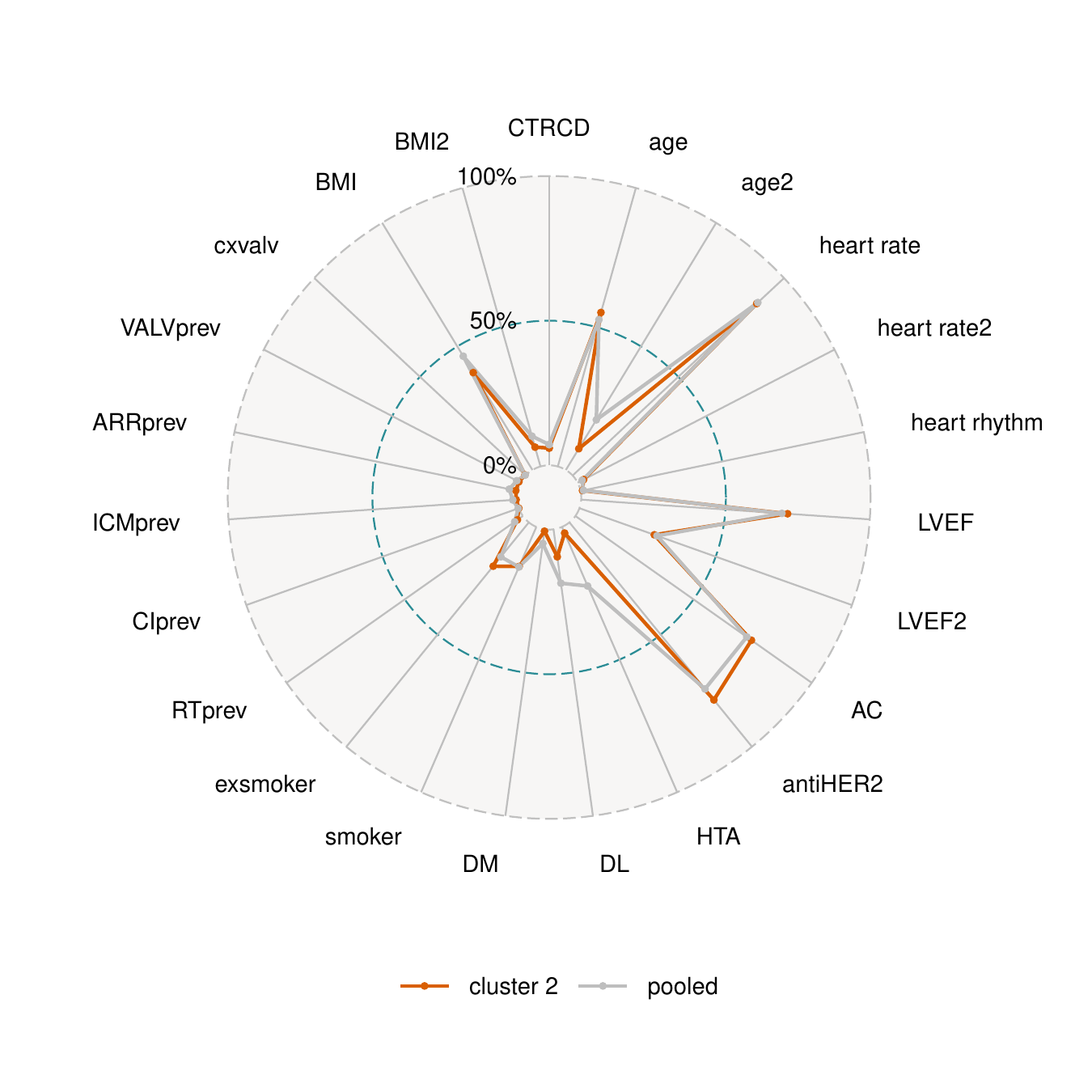}
		\end{subfigure}
		\caption{Breast cancer data. Spider plots for the comparison of the empirical marginal distribution of each variable between estimated cluster (colored line) and pooled dataset (gray line). Left and right panels corresponding to clusters 1 and 2 respectively.}
		\label{application:spider}
	\end{figure}
\end{center}

We summarize the clustering structure learned by our model by building an $(n,n)$ posterior similarity matrix; see Equation \eqref{eq:post:prob:same:cluster}.
From the latter, we recover a point estimate of the clustering based on the minimum posterior expectation of VI \citep{wade:2018}; see also Section \ref{sec:posterior:summaries}.
As the result, we obtain two clusters, that we label as $\widehat{\boldsymbol{c}}_1$ and $\widehat{\boldsymbol{c}}_2$, whose sizes are $n_1 = 101$ and $n_2 = 303$ respectively.
The posterior similarity matrix is represented as a heatmap in Figure \ref{application:psm}, with individuals arranged according to the estimated clusters, specifically those assigned to $\widehat{\boldsymbol{c}}_1$ first and then those in $\widehat{\boldsymbol{c}}_2$.
The two-cluster structure is pretty evident from the matrix since the probabilities of membership to the same group approach value one (zero) for individuals assigned to the same (to a different) estimated cluster.
%
%
%
We then investigate differences between the estimated clusters by comparing the empirical (marginal) distribution of each variable across $\widehat{\boldsymbol{c}}_1$ and $\widehat{\boldsymbol{c}}_2$. For each cluster, we provide a graphical representation based on a spider plot, which includes for each categorical (binary) variable the proportion (percentage values) of observations corresponding to level labeled as $1$ of the variable. These values are reported as colored points joined by lines in each graph; see Figure \ref{application:spider}. Additionally, each plot includes the same proportions as obtained from the pooled sample, namely when no clustering is considered, that are instead represented with grey dots joined by grey lines.
While for cluster 2 (right-side plot) the cluster-proportions are almost aligned with those computed on the pooled dataset, cluster 1 presents a few peculiarities. In particular, patients included in cluster 1 are in general older, as it appears from the higher frequency associated to variable age 2,
and characterized by a higher BMI index. Additionally, the proportion of patients suffering from hypertension (HTA), dyslipidemia (DL), and diabetes mellitus (DM) is in general higher in comparison with the pooled dataset, than those in cluster 2.
We emphasize that the results above allow to capture differences between estimated clusters that are reflected in the \emph{marginal} (empirical) distribution of the variables.
Importantly however, differences may emerge from the \emph{joint} distribution of the variables, and specifically in the dependence (DAG) structure for patients assigned to different estimated clusters.
To this end, in the next section we provide results relative to structure learning, which is carried out at subject-specific level.

\subsection{Structure learning}
\label{sec:application:structurelearning}
\begin{center}
	\begin{figure}
		\begin{subfigure}{.5\textwidth}
			\centering
			\includegraphics[width=1\linewidth]{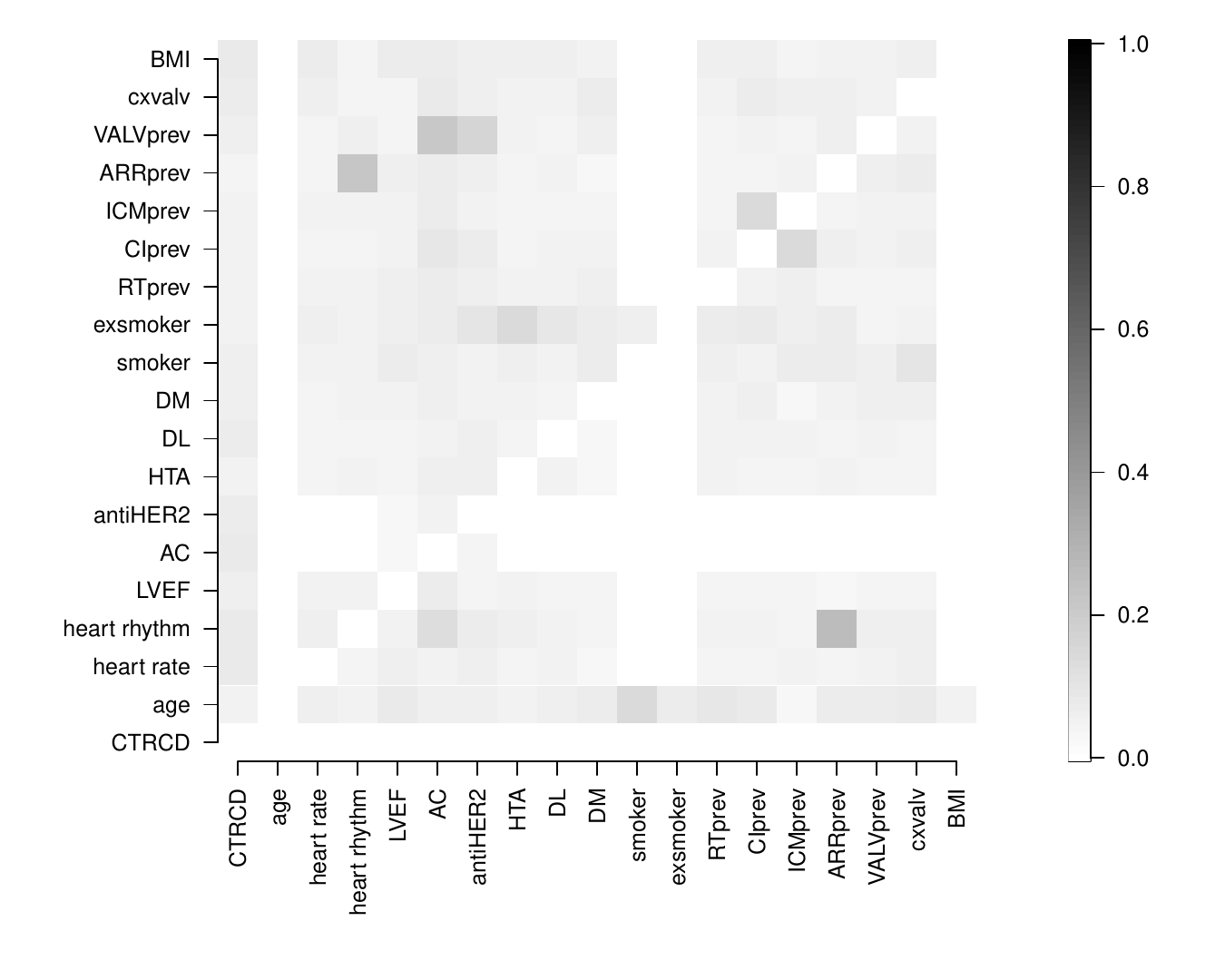}
		\end{subfigure}%
		\begin{subfigure}{.5\textwidth}
			\centering
			\includegraphics[width=1\linewidth]{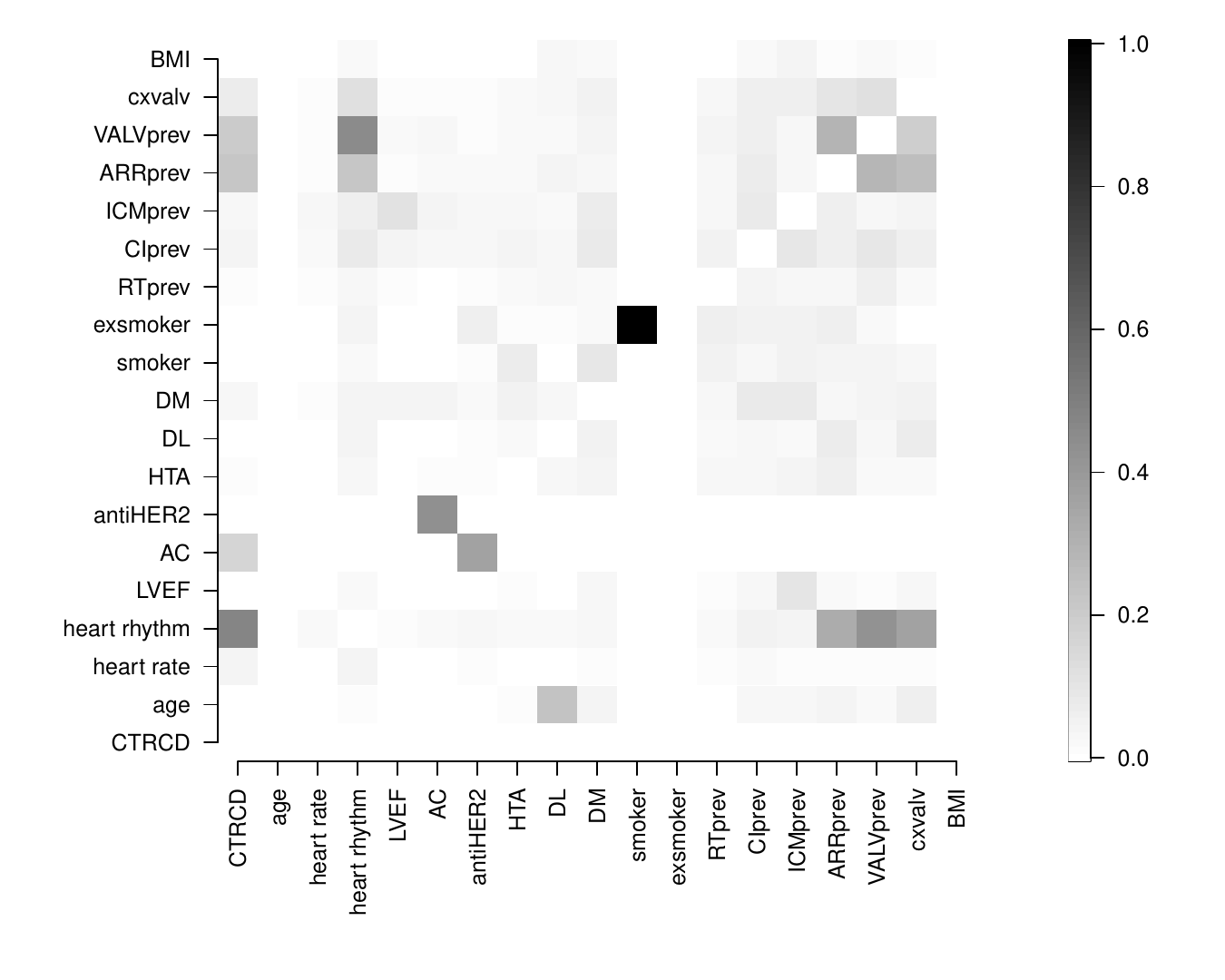}
		\end{subfigure}
		\caption{Breast cancer data. Heat maps of posterior probabilities of edge inclusion for two subject-specific graphs. Left map	corresponds to one subjects in estimated cluster 1; right map to one subject in cluster 2.}
		\label{application:graphs}
	\end{figure}
\end{center}

Following Equation \eqref{eq:post:prob:edge:inclusion}, we provide for each subject $i = 1,\dots,n$ an estimate of the Posterior Probabilities of Inclusion (PPIs), that we collect in a $(q, q)$ matrix.
Because of the structure of our baseline measure, we expect individuals that with high probability are assigned to the same cluster (Figure \ref{application:psm}) to share similar dependence structures, and in particular similar PPIs. By converse, differences in the underlying graphical structure are expected for individuals assigned to distinct estimated clusters.
For two randomly chosen subjects,
whose membership is estimated to be cluster 1 and cluster 2 respectively, the corresponding PPIs are reported as heatmaps in Figure \ref{application:graphs}.
The underlying dependence structures are both characterized by sparsity. This is much more evident in subject from cluster 1, where PPIs are more uniform and there are no edges whose PPI exceeds the $0.5$ threshold.
Differently, the heatmap from cluster 2 shows a few variables that are more strongly related to the outcome CTRCD, specifically AC, together with several risk factors, in particular hearth rhythm, VALVprev, ARRprev.
Accordingly, we expect such differences to imply heterogeneous causal effects between variables. 
We present some of these results in the next section.

\subsection{Causal effects}
\begin{figure*}[t!]
\subfloat{%
	\includegraphics[width=.48\linewidth]{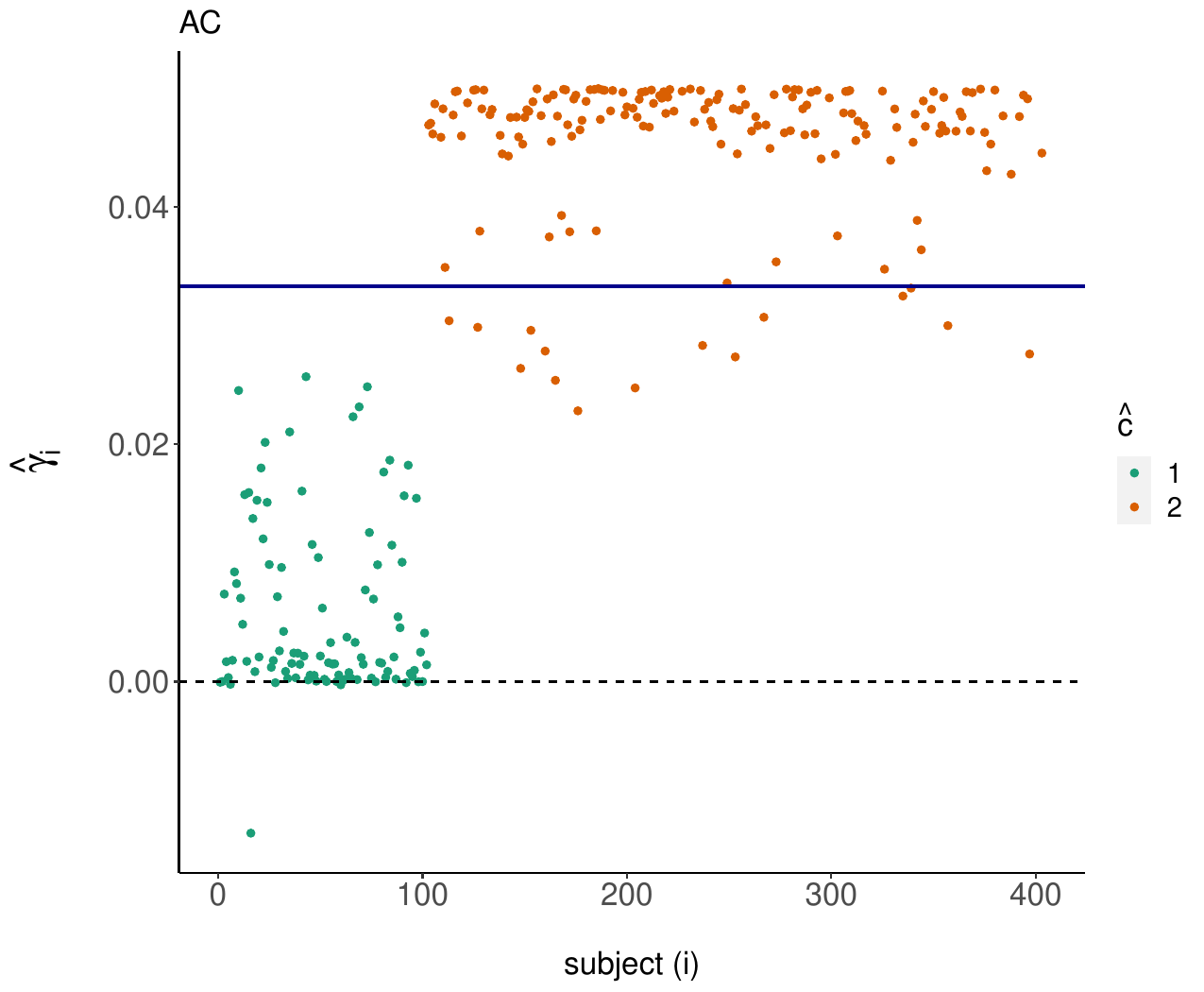}%
}\hfill
{%
	\includegraphics[width=.48\linewidth]{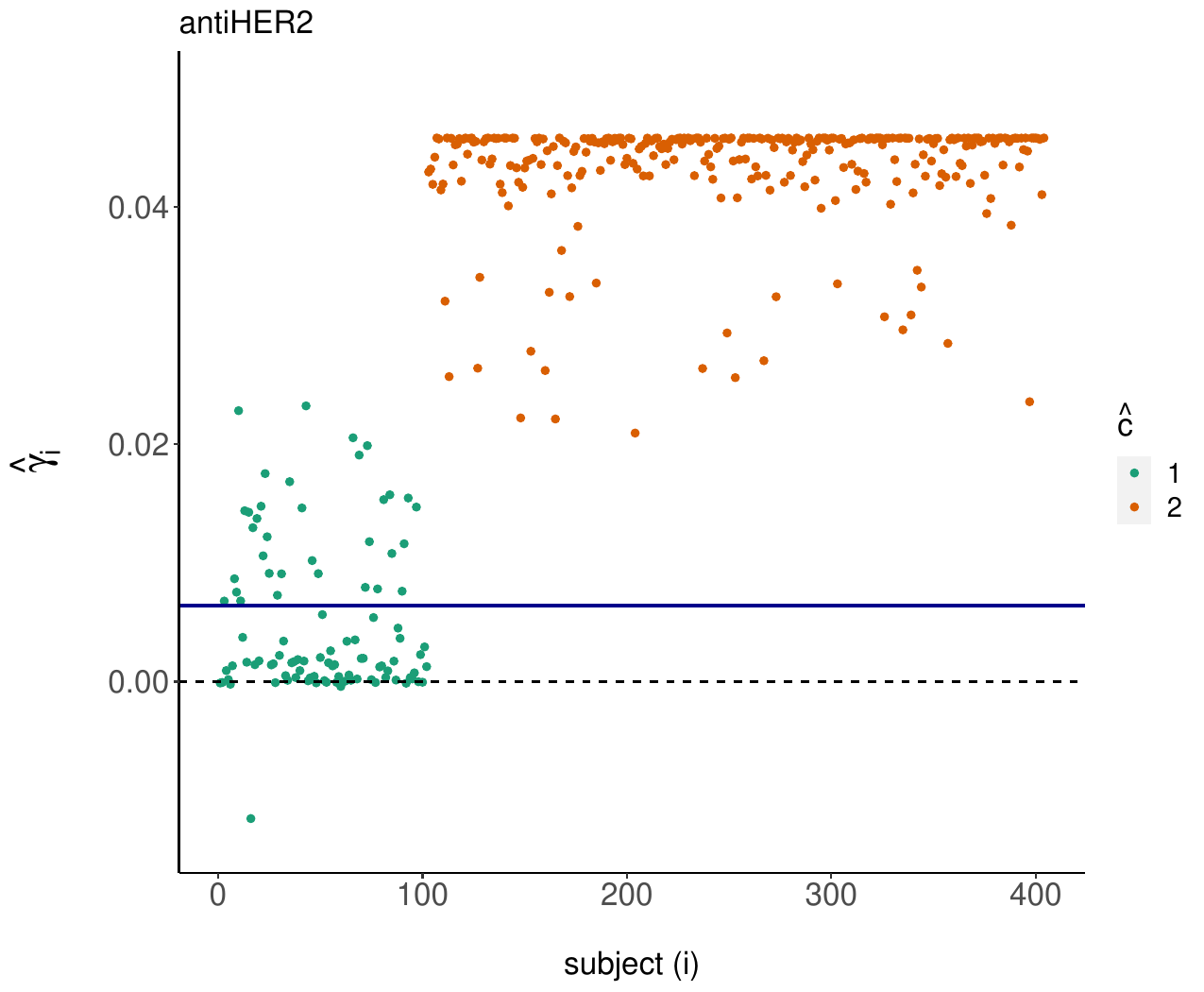}
}%
\caption{Breast cancer data. BMA estimates of subject-specific causal effects, with individuals arranged according to the estimated clustering structure (cluster labeled as $\widehat{\boldsymbol{c}}_1$ in green and $\widehat{\boldsymbol{c}}_2$ in orange). Blue lines correspond to causal-effect estimated obtained under a One-naive strategy disregarding heterogeneity.}
\label{fig:bma:causal:effects}
\end{figure*}

The ultimate goal of our analysis is to quantify the (causal) effect of anticancer therapies on the occurrence of cardiotoxicity.
In particular, patients in the study were treated with therapies based on either antracycline (AC) or trastuzumab (antiHER2).
We then consider the causal effect on the occurrence of cardiotoxicity (variable CTRCD) implied by the administration of AC and antiHER2 therapies.
To this end, we recover from our MCMC output a posterior distribution for each causal-effect parameter and each subject $i=1,\dots,n$, that we summarize through BMA estimates following Equation \eqref{eq:bma}.
%
%
Our results are summarized in the two scatterplots of Figure \ref{fig:bma:causal:effects}, each reporting BMA causal-effect estimates (y-axis) computed across individuals (x-axis) and with subjects arranged according to the estimated clustering with two groups (Section \ref{sec:application:clustering}).
One can appreciate the heterogeneity in the estimates, with individuals assigned to the same cluster sharing similar values, except for a few patients in each group. Interestingly, these subjects are also characterized by a higher uncertainty in cluster allocation between either group 1 or 2; see in particular the posterior similarity matrix in Figure \ref{fig:clustering_sim}.
As an interesting result, AC and antiHER2 treatments in general increase the probability of CTRCD occurrence for individuals assigned to cluster 2, while the effect is less pronounced, or even null, for cluster 1.
Notably, cluster 1 is characterized by older patients and with a higher prevalence of some risk factors; see in particular Figure \ref{application:spider}. Accordingly, in such patients, the occurrence of cardiotoxicity might be due to the presence of such risk factors, that may cause cardiac diseases, rather than implied by the therapy itself. Therefore, the \textit{direct} effect of AC and antiHER2 therapies is lower in comparison with the same estimates in cluster 2.

In addition, to emphasize the role played by population heterogeneity in causal effect estimation, we compare our results with those based on an alternative One-group naive strategy, corresponding to a standard categorical DAG model in which all individuals are assigned to the same cluster.
In such case, causal effect estimates are uniform across subjects and are included as horizontal lines in the two plots.
Each resulting estimate is approximately an average of
cluster-specific causal estimates, suggesting that a causal effect analysis that disregards population heterogeneity would over- and under- estimate the risk of CTRCD development across individuals.

	%
	%
\label{sec:application:causaleffect}

\section{Discussion}
\label{sec:discussion} 

We proposed a modeling framework for structure learning and causal inference under heterogeneity that applies to multivariate categorical data.
Our methodology is based on a Dirichlet Process (DP) mixture of categorical Directed Acyclic Graphs (DAGs) which allows to cluster subjects characterized by similar patterns of dependencies into homogeneous groups, and to provide estimates of causal effects at personalized, namely subject-specific, level.
When adopted for clustering purposes, our method clearly outperforms benchmark strategies that do not account for dependence relations in the joint distribution of variables.
Most importantly, our causal-effect analysis shows that approaches neglecting possible population heterogeneity can provide misleading estimates of causal effects. With regard to our application to breast cancer data, the probability of cardiotoxic side-effects implied by anticancer therapies could be underestimated for several patients, with serious consequences in clinical decision making for optimal therapies' administration.



 
A possible extension of our model could be the analysis of \emph{mixed} multivariate data, which comprise both quantitative and categorical measurements. Specifically in a biomedical setting, one can collect besides categorical clinical features the expression levels of genes that are involved in the progression of the disease. Causal inference methods that integrate such biological information can provide a more precise quantification of causal effects for the development of personalized therapies. 
More in general, multivariate models that can manage mixed data would be valuable for clustering purposes too, as several real-world applications frequently involve data of various types.
Mixed-data represent an interesting framework for graphical modelling which has been addressed in a few works, although without accounting for population heterogeneity; see for instance \citet{Castelletti:2024:Psychometrika}.

The breast cancer dataset provided by \cite{BreastCancerData}
includes Tissue Doppler Imaging (TDI) data, which measure the rate of contraction and relaxation of the cardiac muscle. TDI measurements can be treated as functional data, which whenever included in our analysis could help identifying sub-groups of patients who experienced different side effects of anti-HER2 therapies depending on the presence of cardiac dysfunctions. The inclusion of functional variables in our framework presents challenges that are related to the development of a DAG-based model. As a starting point, \cite{Qiao:et:al:2019} propose a frequentist method to learn dependencies across functional variables, which is based on undirected graphical models and lasso-type penalization techniques.





\bigskip
\begin{center}
	{\large\bf SUPPLEMENTARY MATERIAL}
\end{center}

Supplementary Materials comprise theoretical results on the computation of prior and posterior predictive distributions, details on sampling from the baseline over DAGs, and data generation for our simulation studies.

\black

\bibliographystyle{Chicago}

\bibliography{Bibliography-MM-MC}

\begin{thebibliography}{}

\bibitem[\protect\citeauthoryear{Argiento and De~Iorio}{Argiento and
  De~Iorio}{2022}]{Argiento:De:Iorio:2022}
Argiento, R. and M.~De~Iorio (2022).
\newblock {Is Infinity that Far? A {B}ayesian Nonparametric Perspective of
  Finite Mixture Models}.
\newblock {\em The Annals of Statistics\/}~{\em 50\/}(5), 2641 -- 2663.

\bibitem[\protect\citeauthoryear{Argiento, Filippi-Mazzola, and Paci}{Argiento
  et~al.}{2022}]{argientopaci}
Argiento, R., E.~Filippi-Mazzola, and L.~Paci (2022).
\newblock {Model-based Clustering of Categorical Data based on the Hamming
  Distance}.
\newblock {\em arXiv preprint\/}.

\bibitem[\protect\citeauthoryear{Athey and Imbens}{Athey and
  Imbens}{2016}]{Imbens:et:al:2020}
Athey, S. and G.~Imbens (2016).
\newblock {Recursive Partitioning for Heterogeneous Causal Effects}.
\newblock {\em Proceedings of the National Academy of Sciences\/}~{\em 113},
  965 -- 2020.

\bibitem[\protect\citeauthoryear{Bargagli~Stoffi, De~Witte, and
  Gnecco}{Bargagli~Stoffi et~al.}{2022}]{Bargagli:Stoffi:et:al:2022}
Bargagli~Stoffi, F., K.~De~Witte, and G.~Gnecco (2022).
\newblock {Heterogeneous Causal Effects with Imperfect Compliance: A {B}ayesian
  Machine Learning Approach}.
\newblock {\em The Annals of Applied Statistics\/}~{\em 16\/}(3), 1986 -- 2009.

\bibitem[\protect\citeauthoryear{Bowles, Wellman, Feigelson, Onitilo, Freedman,
  Delate, Allen, Nekhlyudov, Goddard, Davis, Habel, Yood, Mccarty, Magid,
  Wagner, and Team}{Bowles et~al.}{2012}]{med4}
Bowles, E. J.~A., R.~Wellman, H.~S. Feigelson, A.~A. Onitilo, A.~N. Freedman,
  T.~Delate, L.~A. Allen, L.~Nekhlyudov, K.~A.~B. Goddard, R.~L. Davis, L.~A.
  Habel, M.~U. Yood, C.~Mccarty, D.~J. Magid, E.~H. Wagner, and P.~S. Team
  (2012, 09).
\newblock {Risk of Heart Failure in Breast Cancer Patients After Anthracycline
  and Trastuzumab Treatment: A Retrospective Cohort Study}.
\newblock {\em JNCI: Journal of the National Cancer Institute\/}~{\em
  104\/}(17), 1293 -- 1305.

\bibitem[\protect\citeauthoryear{Castelletti}{Castelletti}{2024}]{Castelletti:2024:Psychometrika}
Castelletti, F. (2024).
\newblock {Learning {B}ayesian Networks: A Copula Approach for Mixed-Type
  Data}.
\newblock {\em Psychometrika\/}~{\em 89}, 658 -- 686.

\bibitem[\protect\citeauthoryear{Castelletti and Consonni}{Castelletti and
  Consonni}{2023}]{castellettigaussian}
Castelletti, F. and G.~Consonni (2023).
\newblock {Bayesian Graphical Modeling for Heterogeneous Causal Effects}.
\newblock {\em Statistics in Medicine\/}~{\em 42}, 15--32.

\bibitem[\protect\citeauthoryear{Castelletti, Consonni, and
  Della Vedova}{Castelletti et~al.}{2024}]{castelletti2023joint}
Castelletti, F., G.~Consonni, and M.~L. Della Vedova (2024, 07).
\newblock {Joint Structure Learning and Causal Effect Estimation for
  Categorical Graphical Models}.
\newblock {\em Biometrics\/}~{\em 80\/}(3).

\bibitem[\protect\citeauthoryear{Dempke, Zielinski, Winkler, Silberman,
  Reuther, and Priebe}{Dempke et~al.}{2023}]{med3}
Dempke, W.~C., R.~Zielinski, C.~Winkler, S.~Silberman, S.~Reuther, and
  W.~Priebe (2023).
\newblock {Anthracycline-Induced Cardiotoxicity -- Are we About to Clear this
  Hurdle?}
\newblock {\em European Journal of Cancer\/}~{\em 185}, 94 -- 104.

\bibitem[\protect\citeauthoryear{Dempsey, Rosenthal, Dabas, Kropotova, Lippman,
  and Bishopric}{Dempsey et~al.}{2021}]{med5}
Dempsey, N., A.~Rosenthal, N.~Dabas, Y.~Kropotova, M.~Lippman, and N.~H.
  Bishopric (2021, 07).
\newblock {Trastuzumab-Induced Cardiotoxicity: A Review of Clinical Risk
  Factors, Pharmacologic Prevention, and Cardiotoxicity of other HER2-Directed
  Therapies}.
\newblock {\em Breast Cancer Research and Treatment\/}~{\em 188\/}(17), 21 --
  36.

\bibitem[\protect\citeauthoryear{Dominici, Bargagli~Stoffi, and
  Mealli}{Dominici et~al.}{2021}]{Dominici:Mealli:2021}
Dominici, F., F.~J. Bargagli~Stoffi, and F.~Mealli (2021).
\newblock {From Controlled to Undisciplined Data: Estimating Causal Effects in
  the Era of Data Science Using a Potential Outcome Framework}.
\newblock {\em Harvard Data Science Review\/}~{\em 3\/}(3).

\bibitem[\protect\citeauthoryear{Escobar and West}{Escobar and
  West}{1994}]{escobar}
Escobar, M. and M.~West (1994).
\newblock {Bayesian Density Estimation and Inference Using Mixtures}.
\newblock {\em Journal of the American Statistical Association\/}~{\em
  90\/}(430), 577 -- 588.

\bibitem[\protect\citeauthoryear{Ferguson}{Ferguson}{1973}]{Ferguson:1973}
Ferguson, T.~S. (1973).
\newblock {A Bayesian Analysis of Some Nonparametric Problems}.
\newblock {\em The Annals of Statistics\/}~{\em 1\/}(2), 209 -- 230.

\bibitem[\protect\citeauthoryear{Fr{\"u}hwirth-Schnatter, Malsiner-Walli, and
  Gr{\"u}n}{Fr{\"u}hwirth-Schnatter
  et~al.}{2021}]{Fruhwirth:Schnatter:et:al:2021}
Fr{\"u}hwirth-Schnatter, S., G.~Malsiner-Walli, and B.~Gr{\"u}n (2021).
\newblock {Generalized Mixtures of Finite Mixtures and Telescoping Sampling}.
\newblock {\em Bayesian Analysis\/}~{\em 16\/}(4), 1279 -- 1307.

\bibitem[\protect\citeauthoryear{Geiger and Heckerman}{Geiger and
  Heckerman}{1997}]{g&l}
Geiger, D. and D.~Heckerman (1997, 02).
\newblock {A Characterization of the {D}irichlet Distribution through Global
  and Local Parameter Independence}.
\newblock {\em Annals of Statistics\/}~{\em 25}, 1344 -- 1369.

\bibitem[\protect\citeauthoryear{Goodman}{Goodman}{1974}]{goodman}
Goodman, L.~A. (1974).
\newblock {Exploratory Latent Structure Analysis Using Both Identifiable and
  Unidentifiable Models}.
\newblock {\em Biometrika\/}~{\em 61\/}(2), 215 -- 231.

\bibitem[\protect\citeauthoryear{Hahn, Murray, and Carvalho}{Hahn
  et~al.}{2020}]{Hahn:et:al:2020}
Hahn, P.~R., J.~S. Murray, and C.~M. Carvalho (2020).
\newblock {Bayesian Regression Tree Models for Causal Inference:
  Regularization, Confounding, and Heterogeneous Effects (with Discussion)}.
\newblock {\em Bayesian Analysis\/}~{\em 15\/}(3), 965 -- 2020.

\bibitem[\protect\citeauthoryear{Heckerman, Geiger, and Chickering}{Heckerman
  et~al.}{1995}]{Heckerman:et:al:1995}
Heckerman, D., D.~Geiger, and D.~M. Chickering (1995).
\newblock {Learning {B}ayesian Networks: The Combination of Knowledge and
  Statistical Data}.
\newblock {\em Machine Learning\/}~{\em 20\/}(3), 197 -- 243.

\bibitem[\protect\citeauthoryear{Huang}{Huang}{1998}]{kmodes}
Huang, Z. (1998).
\newblock {Extensions to the k-Means Algorithm for Clustering Large Data Sets
  with Categorical Values}.
\newblock {\em Data Mining and Knowledge Discovery\/}~{\em 2}, 283 -- 304.

\bibitem[\protect\citeauthoryear{Katzorke, Kathrin~Rack, Haeberle,
  Katharina~Neugebauer, Anna~Melcher, Hagenbeck, Forstbauer, Ulrich~Ulmer,
  Soeling, Kreienberg, Fehm, Schneeweiss, Beckmann, Fasching, and
  Janni}{Katzorke et~al.}{2013}]{med2}
Katzorke, N., B.~Kathrin~Rack, L.~Haeberle, J.~Katharina~Neugebauer,
  C.~Anna~Melcher, C.~Hagenbeck, H.~Forstbauer, H.~Ulrich~Ulmer, U.~Soeling,
  R.~Kreienberg, T.~N. Fehm, A.~Schneeweiss, M.~W. Beckmann, P.~A. Fasching,
  and W.~Janni (2013).
\newblock {Prognostic Value of HER2 on Breast Cancer Survival}.
\newblock {\em Journal of Clinical Oncology\/}~{\em 31\/}(15), 600 -- 640.

\bibitem[\protect\citeauthoryear{Lee, Bargagli~Stoffi, and Dominici}{Lee
  et~al.}{2021}]{Bargagli:Stoffi:2021:CRE}
Lee, K., F.~Bargagli~Stoffi, and F.~Dominici (2021).
\newblock {Causal Rule Ensemble: Interpretable Inference of Heterogeneous
  Treatment Effects}.
\newblock {\em arXiv preprint\/}.

\bibitem[\protect\citeauthoryear{Linero and Antonelli}{Linero and
  Antonelli}{2022}]{Linero:Antonelli:2022}
Linero, A.~R. and J.~L. Antonelli (2022).
\newblock {The How and Why of Bayesian Nonparametric Causal Inference}.
\newblock {\em Wiley Interdisciplinary Reviews: Computational
  Statistics\/}~{\em 15}.

\bibitem[\protect\citeauthoryear{Linzer and Lewis}{Linzer and
  Lewis}{2011}]{linzer}
Linzer, D.~A. and J.~B. Lewis (2011).
\newblock {poLCA: An R Package for Polytomous Variable Latent Class Analysis}.
\newblock {\em Journal of Statistical Software\/}~{\em 42\/}(10), 1 -- 29.

\bibitem[\protect\citeauthoryear{Lotrionte, Biondi-Zoccai, Abbate, Lanzetta,
  D'Ascenzo, Malavasi, Peruzzi, Frati, and Palazzoni}{Lotrionte
  et~al.}{2013}]{med6}
Lotrionte, M., G.~Biondi-Zoccai, A.~Abbate, G.~Lanzetta, F.~D'Ascenzo,
  V.~Malavasi, M.~Peruzzi, G.~Frati, and G.~Palazzoni (2013).
\newblock {Review and Meta-Analysis of Incidence and Clinical Predictors of
  Anthracycline Cardiotoxicity}.
\newblock {\em The American Journal of Cardiology\/}~{\em 112\/}(12), 1980 --
  1984.

\bibitem[\protect\citeauthoryear{Ma, Hobbs, and Stingo}{Ma
  et~al.}{2015}]{Stingo:et:al:2015}
Ma, J., B.~Hobbs, and F.~Stingo (2015).
\newblock {Statistical Methods for Establishing Personalized Treatment Rules in
  Oncology}.
\newblock {\em BioMed Research International\/}~{\em 2015\/}(1), 670691.

\bibitem[\protect\citeauthoryear{MacQueen}{MacQueen}{1967}]{MacQueen:1967}
MacQueen, J. (1967).
\newblock Some methods for classification and analysis of multivariate
  observations.
\newblock In {\em Volume 1: Statistics}, Proceedings of the Fifth Berkeley
  Symposium on Mathematical Statistics and Probability, pp.\  281 -- 297.
  University of California Press.

\bibitem[\protect\citeauthoryear{Malsiner-Walli, Grün, and
  Frühwirth-Schnatter}{Malsiner-Walli et~al.}{2024}]{Malsiner:et:al:2024}
Malsiner-Walli, G., B.~Grün, and S.~Frühwirth-Schnatter (2024).
\newblock {Without Pain -- Clustering Categorical Data Using a Bayesian Mixture
  of Finite Mixtures of Latent Class Analysis Models}.
\newblock {\em arXiv preprint\/}.

\bibitem[\protect\citeauthoryear{M{\"u}ller and Rodriguez}{M{\"u}ller and
  Rodriguez}{2013}]{muller2013dirichlet}
M{\"u}ller, P. and A.~Rodriguez (2013).
\newblock Dirichlet process.
\newblock In {\em {Nonparametric Bayesian Inference}}, Volume~9, pp.\  23 --
  42. Institute of Mathematical Statistics.

\bibitem[\protect\citeauthoryear{Neal}{Neal}{2000}]{Neal:2000}
Neal, R.~M. (2000).
\newblock {Markov Chain Sampling Methods for {D}irichlet Process Mixture
  Models}.
\newblock {\em Journal of Computational and Graphical Statistics\/}~{\em
  9\/}(2), 249 -- 265.

\bibitem[\protect\citeauthoryear{Oganisian, Mitra, and Roy}{Oganisian
  et~al.}{2021}]{Oganisian:et:al:2021}
Oganisian, A., N.~Mitra, and J.~A. Roy (2021).
\newblock {A Bayesian Nonparametric Model for Zero-Inflated Outcomes:
  Prediction, Clustering, and Causal Estimation}.
\newblock {\em Biometrics\/}~{\em 77\/}(1), 125 -- 135.

\bibitem[\protect\citeauthoryear{Pearl}{Pearl}{2000}]{Pearl:2000}
Pearl, J. (2000).
\newblock {\em {Causality: Models, Reasoning, and Inference}}.
\newblock Cambridge University Press, Cambridge.

\bibitem[\protect\citeauthoryear{Pearl}{Pearl}{2003}]{pearl1}
Pearl, J. (2003).
\newblock {Statistics and Causal Inference: A Review}.
\newblock {\em Sociedad de Estadistica e Investigacion Operativa\/}~{\em 12},
  101--165.

\bibitem[\protect\citeauthoryear{Pi{\~n}eiro-Lamas, L{\'o}pez-Cheda, Cao,
  Ramos-Alonso, Gonz{\'a}lez-Barbeito, Barbeito-Caama{\~n}o, and
  Bouzas-Mosquera}{Pi{\~n}eiro-Lamas et~al.}{2023}]{BreastCancerData}
Pi{\~n}eiro-Lamas, B., A.~L{\'o}pez-Cheda, R.~Cao, L.~Ramos-Alonso,
  G.~Gonz{\'a}lez-Barbeito, C.~Barbeito-Caama{\~n}o, and A.~Bouzas-Mosquera
  (2023).
\newblock {A Cardiotoxicity Dataset for Breast Cancer Patients}.
\newblock {\em Scientific Data\/}~{\em 10\/}(1), 527.

\bibitem[\protect\citeauthoryear{Qiao, Guo, and James}{Qiao
  et~al.}{2019}]{Qiao:et:al:2019}
Qiao, X., S.~Guo, and G.~M. James (2019).
\newblock {Functional Graphical Models}.
\newblock {\em Journal of the American Statistical Association\/}~{\em 525},
  211 -- 222.

\bibitem[\protect\citeauthoryear{Rodr{\'i}guez, Lenkoski, and
  Dobra}{Rodr{\'i}guez et~al.}{2011}]{Rodriguez:et:al:2009}
Rodr{\'i}guez, A., A.~Lenkoski, and A.~Dobra (2011).
\newblock {Sparse Covariance Estimation in Heterogeneous Samples}.
\newblock {\em Electronic Journal of Statistics\/}~{\em 5}, 981 -- 1014.

\bibitem[\protect\citeauthoryear{Roy, Lum, and Daniels}{Roy
  et~al.}{2016}]{Roy:et:al:2016}
Roy, J., K.~J. Lum, and M.~J. Daniels (2016).
\newblock {A Bayesian Nonparametric Approach to Marginal Structural Models for
  Point Treatments and a Continuous or Survival Outcome}.
\newblock {\em Biostatistics\/}~{\em 18\/}(1), 32 -- 47.

\bibitem[\protect\citeauthoryear{Rubin}{Rubin}{2005}]{Rubin:2005}
Rubin, D.~B. (2005).
\newblock {Causal Inference Using Potential Outcomes: Design, Modeling,
  Decisions}.
\newblock {\em Journal of the American Statistical Association\/}~{\em
  100\/}(469), 322 -- 331.

\bibitem[\protect\citeauthoryear{Sethuraman}{Sethuraman}{1994}]{Sethuraman:1994}
Sethuraman, J. (1994).
\newblock {A Constructive Definition of {D}irichlet Priors}.
\newblock {\em Statistica Sinica\/}~{\em 4\/}(2), 639 -- 650.

\bibitem[\protect\citeauthoryear{Slamon, Clark, Wong, Levin, Ullrich, and
  McGuire}{Slamon et~al.}{1987}]{med1}
Slamon, D.~J., G.~M. Clark, S.~G. Wong, W.~J. Levin, A.~Ullrich, and W.~L.
  McGuire (1987).
\newblock {Human Breast Cancer: Correlation of Relapse and Survival with
  Amplification of the {HER}-2/Neu Oncogene}.
\newblock {\em Science\/}~{\em 235\/}(4785), 177 -- 182.

\bibitem[\protect\citeauthoryear{Wade and Ghahramani}{Wade and
  Ghahramani}{2018}]{wade:2018}
Wade, S. and Z.~Ghahramani (2018).
\newblock {Bayesian Cluster Analysis: Point Estimation and Credible Balls (with
  Discussion)}.
\newblock {\em Bayesian Analysis\/}~{\em 13\/}(2), 559 -- 626.

\bibitem[\protect\citeauthoryear{Zorzetto, Bargagli~Stoffi, Canale, and
  Dominici}{Zorzetto et~al.}{2024}]{Zorzetto:Canale:et:al:2024}
Zorzetto, D., F.~Bargagli~Stoffi, A.~Canale, and F.~Dominici (2024).
\newblock {Confounder-Dependent Bayesian Mixture Model: Characterizing
  Heterogeneity of Causal Effects in Air Pollution Epidemiology}.
\newblock {\em Biometrics\/}~{\em 80\/}(2).

\end{thebibliography}


\begin{thebibliography}{}

\bibitem[\protect\citeauthoryear{Castelletti, Consonni, and
  Della Vedova}{Castelletti et~al.}{2024}]{castelletti2023joint}
Castelletti, F., G.~Consonni, and M.~L. Della Vedova (2024, 07).
\newblock {Joint Structure Learning and Causal Effect Estimation for
  Categorical Graphical Models}.
\newblock {\em Biometrics\/}~{\em 80\/}(3).

\end{thebibliography}

\end{document}


\doublespacing
	
	\maketitle


	\section{Details on the DAG marginal likelihood}
	
	Consider a DAG $\D$ and an $(n, q)$ data matrix $\bX$ collecting $n$ $q$-variate categorical observations $\bx^{(i)}=(x_1^{(i)},\dots,x_q^{(i)})^\top, i=1,\dots,n$. Assume the likelihood function $p(\bX\g\btheta,\D)$ and prior $p(\btheta\g\D)$ as in Section 1 and 3.1 of the main paper respectively. Then, the marginal likelihood of a categorical DAG model is given by
	\begin{equation}
		\label{eq:marginal:likelihood}
		m(\bX \g \D) = \int p(\bX \g \btheta, \D) \ p(\btheta \g \D) \ d\btheta = \\
		\prod_{j = 1}^q \left\{ \prod_{s \in \Xpaj} \frac{h\big(\aj\big)}{h\big(\aj + \nj\big)} \right\}, 
	\end{equation}
	where in particular
	\begin{align*}
		h\big(\aj\big) &= \frac{\Gamma\Big(\sum_{m \in \Xj} \ajpaj\Big)}{\prod_{m \in \Xj} \Gamma\Big(\ajpaj\Big)},
		\\
		h\big(\aj + \nj\big) &= \frac{\Gamma\Big(\sum_{m \in \Xj} \ajpaj + \njms\Big)}{\prod_{m \in \Xj} \Gamma\Big(\ajpaj + \njms\Big)}
	\end{align*}
	are the prior and posterior normalizing constants respectively. See  \cite{castelletti2023joint} for details.
	Moreover, under the default choice $\ajpaj = a/|\Xj|$, Equation \eqref{eq:marginal:likelihood} reduces to
	\begin{equation}
		\begin{aligned}
			\label{eq:ml:default}
			m(\bX \g \D) = \prod_{j = 1}^q \left\{ \prod_{s \in \Xpaj} \frac{\Gamma \big(a/ |\Xpaj| \ \big)}{\Gamma \Big(a/ |\Xpaj| + \npajs \ \Big)} 
			\prod_{m \in \Xj}  \frac{ \Gamma \Big(a/ |\Xfaj| + \njms \ \Big)}{\Gamma \big( a/ |\Xfaj| \big)} \right\}.
		\end{aligned}
	\end{equation}
	
	\vspace{0.2cm}
	
	\section{Proposal distribution for DAGs and sampling from the baseline over the DAG space}
	
	We provide details about the construction of a proposal distribution to explore the space of DAGs. This is required by our MCMC sampler to update DAGs through a Metropolis Hastings step and to sample from the baseline over DAGs in the update of cluster indicators; see in particular Section 4.1 of our paper.
	
	\vspace{0.3cm}
	
	Consider $\D\in\mathcal{S}_q$ where $\mathcal{S}_q$ is a collection of DAGs, such as the space of \textit{all} DAGs having $q$ nodes. To update $\D$, we first draw a new DAG $\widetilde{\D}$ from a proposal distribution which is based on three types of operators that locally modify $\D$: insert a directed edge (InsertD $u \rightarrow v$ for short), delete a directed edge (DeleteD $u\rightarrow v$) and reverse a directed edge (ReverseD $u \rightarrow v$). We then construct the set of valid operators $\mathcal{O}_{\D}$, that is operators whose resulting graph is in $\mathcal{S}_q$.
	Finally, we propose $\widetilde{\D}$ by uniformly sampling an element in $\mathcal{O}_{\D}$ and applying it to $\D$. Since there is a one-to-one correspondence between each operator and the resulting DAG, the probability of transition is $q(\widetilde{\D}\g\D)=1/|\mathcal{O}_{\D}|$, for each $\widetilde{\D}$ direct successor of $\D$.
	
	\vspace{0.3cm}
	
	Since the enumeration of all DAGs in $\mathcal{S}_q$ is unfeasible in practice, direct sampling from the baseline $p(\D)$ can be performed through an acceptance-rejection method.
	For a given DAG $\D$, let $N(\D)$ be the set of all its direct successors, each one obtained by applying an operator in the set $\mathcal{O}_{\D}$ defined above.
	We first uniformly sample a DAG $\widetilde{\D}$ from $N(\D)$, which occurs with probability $q(\widetilde{\D}\g\D)=1/|N(\D)|$, for each $\widetilde{\D}\in N(\D)$. Hence, we move to $\widetilde{\D}$ with probability
	\be
	\alpha_{\widetilde{\D}}=
	\min
	\left\{
	1; \frac{p(\widetilde{\D})}{p(\D)}\cdot \frac{q(\D\g\widetilde{\D})}{q(\widetilde{\D}\g\D)}
	\right\}.
	\ee
	Importantly, to compute $\alpha_{\widetilde{\D}}$ we only need to evaluate the \emph{ratio} of the priors
	$p(\widetilde{\D})/p(\D)=r$, which avoids the computation of normalizing constants over the space of DAGs; see also Section 3.2 of the main paper.
	Moreover, the ratio of the two proposal probabilities reduces to
	$q(\D\g\widetilde{\D})/q(\widetilde{\D}\g\D)=|\mathcal{O}_{\D}|/|\mathcal{O}_{\widetilde{\D}}|$
	which requires the enumeration of all the direct successors of $\D$ and $\widetilde{\D}$. While this is feasible with a relatively small computational cost, it was observed empirically that the approximation
	$q(\D\g\widetilde{\D})/q(\widetilde{\D}\g\D)\approx 1$
	does not produce a relevant loss in terms of accuracy.

	\section{Proofs of Propositions 4.1 and 4.2}
	
	In this section we provide the proofs of Propositions 4.1 and 4.2 on posterior predictive distributions, whose statements are included in Section 4.2 of our paper.
	
	\subsection{Proof of Proposition 4.1}
	
	\begin{customprop}{4.1}[Posterior predictive - non-empty cluster] \label{post_pred_nonempty}
		For a given cluster $k$, consider the data matrix $\bX^{(k)}$ collecting the $n_k$ observations $\big\{\bx^{(l)} : \xi_l=k\big\}$ and an observation $\bx^{(i)}$.
		Then, the posterior predictive of $\bx^{(i)}$ given $\big\{\bx^{(l)} : l \neq i, \xi_l=k\big\}$ under DAG $\D_k$ is
		\begin{equation} \label{posteriorpredictive}
			\begin{aligned}
				p(\xii \g \{\xl : l \neq i, \xi_l = k\}, \D_k) =
				\prod_{j = 1}^q \left\{ \frac{a/|\Xfaj| + \nfaj - \mathbbm{1}\{\xi_i=k\}}{ a/|\Xpaj| + \npaj - \mathbbm{1}\{\xi_i=k\}} \right\} 
			\end{aligned}
		\end{equation}
		where $\mj=\xii_j, \sj=\xii_{\pa(j)}$ and
		$$
		\nfaj=\sum_{l : \xi_l = k} \mathbbm{1}\big\{\xl_{\fa(j)} = (\mj, \sj) \big\}, \quad \npaj = \sum_{l : \xi_l = k} \mathbbm{1}\big\{\xl_{\pa(j)} = \sj\big\}.
		$$
	\end{customprop}
	\textit{Proof.}
	We first write the posterior predictive as the ratio of two marginal likelihoods, defined as in \eqref{eq:ml:default}, 
	\begin{equation} \label{eq:postpred}
		\begin{aligned}
			p\big(\xii \g \{\xl : l \neq i, \xi_l = k\}, \D_k\big) = \frac{m \big(\xii,  \xnoik \g \D_k\big)}{m \big(\xnoik \g \D_k\big)},
		\end{aligned}
	\end{equation}
	where $\xnoik = \big\{\xl : l \neq i, \xi_l = k\big\}$. Now notice that DAG $\D_k$ is the same across numerator and denominator and accordingly all terms in \eqref{eq:marginal:likelihood} depending on $\D_k$ but not on the data, namely the prior normalizing constants, cancel out. Therefore, we obtain
	\begin{equation*}
		p\big(\xii \g \{\xl : l \neq i, \xi_l = k\}, \D_k\big) = \prod_{j = 1}^q\left\{
		\prod_{s \in \X_{\pa(j)}}
		\frac{h\big(\aj + \nfatc(-i) \big)}{h\big(\aj + \nfatc(+i) \big)}
		\right\},
	\end{equation*}
	where
	$\nfatc(-i)$ is the table of counts for variable $X_j$ \black given level $s \in \Xpaj$ which is obtained from the data matrix $\bX^{(k)}_{-i}$, i.e.~using all the observations assigned to cluster $k$ excluding $\bx^{(i)}$ if $\xi_i = k$. Similarly for $\nfatc(+i)$, which is obtained from $\bX^{(k)}_{+i}$, i.e.~all observations assigned to cluster $k$ including $\bx^{(i)}$ if $i$ is not currently assigned to that cluster, that is if $\xi_i \neq k$.
	Moreover, by writing explicitly the posterior normalizing constants as in \eqref{eq:ml:default}, we obtain
	\begin{equation*}
		\begin{aligned}
			p\big(\xii \g \{\xl &: l \neq i, \xi_l = k\}, \D_k\big) \\
			&=\prod_{j = 1}^q \left\{ \prod_{s \in \Xpaj} \frac{\Gamma \left( a/ |\Xpaj| +\prescript{}{k}n^{\pa(j)}_s (-i)\right)}{\Gamma \left( a/ |\Xpaj| +\prescript{}{k}n^{\pa(j)}_s (+i)\right)} \prod_{m \in \Xj} \frac{\Gamma \left( a/ |\Xpaj| +\prescript{}{k}n^{\fa(j)}_{(m,s)} (+i)\right)}{\Gamma \left( a/ |\Xpaj| +\prescript{}{k}n^{\fa(j)}_{(m,s)} (-i)\right)}\right\}.
		\end{aligned}
	\end{equation*}
	Now notice that the counts involved in the previous expression only differ by the inclusion/removal of subject $\bx^{(i)}$ from the data matrix $\bX^{(k)}$.
	As a consequence, all counts in $\nfatc(+i)$ and $\nfatc(-i)$ are equal, except for those corresponding to the configuration attained by $\xii$, say $\tilde{s}_j = \xii_{\pa(j)}$ and $\tilde{m}_j = \xii_j$; similarly for $\npatc(+i)$ and $\npatc(-i)$. Accordingly, the predictive probability above simplifies to
	\begin{equation} \label{eq:finalexpr_postpred}
		\begin{aligned}
			p\big(\xii \g \{\xl &: l \neq i, \xi_l = k\}, \D_k\big) \\
			&=
			\prod_{j = 1}^q \left\{ \frac{\Gamma \left(a/|\Xpaj| + \npaj(-i) \right)}{\Gamma \left(a/|\Xpaj| + \npaj(+i)\right)} \ \cdot \ \frac{\Gamma \left(a/|\Xfaj| + \nfaj(+i) \right)}{\Gamma \left(a/|\Xfaj| + \nfaj(-i) \right)} \right\}.
		\end{aligned}
	\end{equation}
	We can now distinguish two cases: 
	(i) $\xi_i = k$
	and (ii) $\xi_i \neq k$.
	%
	
	\vspace{0.2cm}
	\noindent In case (i), because $i$ is currently assigned to cluster $k$, we actually need to remove it from $\bX^{(k)}$. Therefore, we have $\npaj(-i) = \npaj -1$ and similarly $\nfaj = \nfaj - 1$, while $\npaj(+i) = \npaj$ and $\nfaj(+i) = \nfaj$. \\
	As a consequence, Equation \eqref{eq:finalexpr_postpred} becomes
	\begin{equation*}
		\begin{aligned}
			p\big(\xii \g \{\xl &: l \neq i, \xi_l = k\}, \D_k\big) \\
			&=\prod_{j = 1}^q \left\{ \frac{\Gamma \left(a/|\Xpaj| + \npaj - 1\right)}{\Gamma \left(a/|\Xpaj| + \npaj\right)} \ \cdot \ \frac{\Gamma \left(a/|\Xfaj| + \nfaj \right)}{\Gamma \left(a/|\Xfaj| + \nfaj - 1 \right)} \right\}.
		\end{aligned}
	\end{equation*}
	Finally, by using the property of Gamma functions $\Gamma(c + 1) = c \Gamma(c)$, for arbitrary $c>0$,
	we obtain
	\begin{equation*}
		p\big(\xii \g \{\xl : l \neq i, \xi_l = k\}, \D_k\big)
		=
		\prod_{j = 1}^q \left\{ \frac{\left(a/|\Xfaj| + \nfaj - 1 \right)}{ \left( a/|\Xpaj| + \npaj - 1 \right)}\right\}.
	\end{equation*}

	\vspace{0.2cm}
	\noindent In case (ii) instead, because individual $i$ is not in cluster $k$, we actually need to include it. Therefore, $\npaj(+i) = \npaj + 1$ and $\nfaj(+i) = \nfaj + 1$, while $\npaj(-i) = \npaj$ and $\nfaj(-i) = \nfaj$. Therefore, by proceeding similarly as in case (i), we obtain 
	\begin{equation*}
		p\big(\xii \g \{\xl : l \neq i, \xi_l = k\}, \D_k\big)
		=\prod_{j = 1}^q \left\{ \frac{\left(a/|\Xfaj| + \nfaj \right)}{ \left( a/|\Xpaj| + \npaj\right)}\right\}.
	\end{equation*}
	Finally,
	we can combine the results under cases (i) and (ii) by writing
	\begin{equation} \label{posteriorpredictive}
		\begin{aligned}
			p(\xii \g \{\xl : l \neq i, \xi_l = k\}, \D_k) =
			\prod_{j = 1}^q \left\{ \frac{a/|\Xfaj| + \nfaj - \mathbbm{1}\{\xi_i=k\}}{ a/|\Xpaj| + \npaj - \mathbbm{1}\{\xi_i=k\}} \right\} 
		\end{aligned}
	\end{equation}
	where $\mathbbm{1}\{\xi_i=k\}=1$ if $\xi_i=k$, $0$ otherwise, and the expression coincides with the statement of the proposition.

	\subsection{Proof of Proposition 4.2}
	
	\begin{customprop}{4.2}[Posterior predictive - empty cluster] \label{post_pred_nonempty}
		For a new cluster $k = K + 1$, the posterior predictive of $\bx^{(i)}$ coincides with the marginal likelihood and is given by
		\begin{equation}
			p(\xii \g \D_{k}) =  \prod_{j = 1}^q \frac{1}{|\X_j|}.
		\end{equation}
	\end{customprop}
	\textit{Proof.} Notice that if $k=K+1$ is an empty cluster containing no observations, the posterior predictive coincides with $p(\xii \g \D_k)$, namely the marginal likelihood of DAG $\D_k$ evaluated at $\xii$ only.
	Accordingly, to all configurations that are different from the the one attained by $\xii$, will be assigned frequencies equal to zero. Therefore, if we let $\tilde{s}_j = \xii_{\pa(j)}$ and $\tilde{m}_j = \xii_j$, the general expression of the marginal likelihood in Equation \eqref{eq:ml:default} reduces to
	\begin{equation}
		\begin{aligned} \label{eq:emptycl}
			p(\xii \g \D_k) = \prod_{j = 1}^q \left\{ \frac{\Gamma(a/ |\Xpaj|)}{\Gamma(a / |\Xpaj| + n^{\pa(j)}_{\tilde{s}_j})}\ \cdot \ \frac{\Gamma(a / |\Xfaj| + n^{\fa(j)}_{(\tilde{m}_j, \tilde{s}_j)})}{\Gamma(a/|\Xfaj|)} \right\}.
		\end{aligned}
	\end{equation}
	Moreover, since the marginal likelihood is based on a sample of size one, $\xii$, we have $n^{\pa(j)}_{\tilde{s}_j}=1$ and $n^{\fa(j)}_{(\tilde{m}_j, \tilde{s}_j)}=1$. Again using the property of Gamma functions for which $\Gamma(c + 1) = c \Gamma(c)$ with $c > 0$, Equation \eqref{eq:emptycl} reduces to
	$$
	p(\xii \g \D_k) = \prod_{j = 1}^q\left\{ \frac{a / |\Xfaj|}{a/ |\Xpaj|}\right\} = \prod_{j = 1}^q \frac{1}{|\Xj|},
	$$
	which coincides with the statement of the proposition.

	\section{Data generation for simulations}
	
	In this section we provide details relative to data generation for our simulation studies (Section 5 of the paper).
	We consider settings with $q = 10$ nodes, number of clusters $K = 2$ and sample sizes $n_1=n_2$ that we range in $\{100, 200, 500\}$. Two DAGs $\D_1$ and $\D_2$ are generated independently under a probability of edge inclusion $\pi=0.2$,
	and two corresponding categorical datasets, $\bX^{(1)}$ and $\bX^{(2)}$, are built by discretization of latent continuous data that we generate from a Gaussian DAG-model as follows.
	Let $Z_1,\dots,Z_q \g \bSigma_k \sim \mathcal{N}_q(\boldsymbol{0}, \bSigma_k)$, where the covariance matrix $\bSigma_k$ can be written in terms of its modified Cholesky decomposition as $\bSigma_k=\bL_k^{-\top}\bD_k\bL_k^{-1}$. Under such reparameterization, $\bD_k$ is a diagonal matrix, while $\bL_k$ has diagonal entries equal to one and off-diagonal elements different from zero if and only if $u\rightarrow v$ is in $\D_k$. We randomly draw the non-null elements of $\bL_k$ uniformly in $[-2, -1] \cup [1, 2]$, while we fix $\bD_k = \bI_q$. Then, $n_k$ i.i.d observations are generated from $\mathcal{N}_q(\boldsymbol{0}, \bSigma_k)$ and collected into an $(n_k,q)$ data matrix $\bZ^{(k)}$ whose $(i,j)$-th element is $\bZ_{ij}$. These continuous data are finally discretized into binary observations as
	\begin{equation}
		\bX^{(k)}_{ij} =
		\begin{cases}
			0 & \text{if} \ \bZ^{(k)}_{ij} < g_j \\
			1 & \text{if} \ \bZ^{(k)}_{ij} \ge g_j
		\end{cases}
	\end{equation}
	where $g_j \in (-\infty,+\infty)$ is a threshold that we generate randomly across variables and clusters. The choice of such thresholds can be crucial for cluster identification because it affects the proportion of zeros and ones in the (marginal) distributions of variables across clusters. We sample $g_j \sim \textnormal{Unif}(\hat{z}_{j, \alpha}, \hat{z}_{j, 1-\alpha})$, where $\hat{z}_{j, \alpha}$ denotes the quantile of order $\alpha$ in the empirical distribution of $Z_j$.
	Notice that when $\alpha$ is close to $0.5$ the marginal distributions of the variables are similar between  groups, so that cluster identification might be more difficult because differences between clusters are possibly due to different conditional independencies in the joint distribution only. By converse, deviations from the $0.5$ value imply in general differences between marginal distributions across groups, which in turn can ease cluster identification; see also our results in Section 5 of the main paper.

	\bibliographystyle{Chicago}
	\bibliography{Bibliography-MM-MC}